\documentclass[prx,aps,amsmath,amssymb,reprint]{revtex4-2}

\usepackage[utf8]{inputenc}
\usepackage[T1]{fontenc}
\usepackage{graphicx}
\usepackage{xcolor}
\usepackage[hidelinks]{hyperref}
\usepackage{listings}

\renewcommand*{\vec}[1]{\mathbf{#1}}
\newcommand{\rv}{{\vec{r}}}

\newcommand{\pv}{{\vec{p}}}

\newcommand{\kB}{k_\mathrm{B}}

\newcommand{\rmexc}{{\rm exc}}
\newcommand{\rmext}{{\rm ext}}
\newcommand{\rmid}{{\rm id}}
\newcommand{\ells}{{\ell'}}
\newcommand{\ellss}{{\ell''}}
\newcommand{\ia}{{i\in{\cal N}_\alpha}}
\newcommand{\alphas}{{\alpha'}}
\newcommand{\alphass}{{\alpha''}}
\newcommand{\all}[1]{{\{#1\}}}

\DeclareMathOperator{\Tr}{Tr}
\DeclareMathOperator{\triv}{triv}
\DeclareMathOperator{\cov}{cov}
\DeclareMathOperator{\var}{var}

\graphicspath{{./figures/}}

\begin{document}

\title{Multivariate hyperdensity functional theory for inhomogeneous equilibrium fluids: From first principles to simulation-based machine learning}

\author{Florian Sammüller}
\email{Florian.Sammueller@uni-bayreuth.de}
\affiliation{Theoretische Physik II, Physikalisches Institut, Universität Bayreuth, D-95447 Bayreuth, Germany}

\author{Matthias Schmidt}
\email{Matthias.Schmidt@uni-bayreuth.de}
\affiliation{Theoretische Physik II, Physikalisches Institut, Universität Bayreuth, D-95447 Bayreuth, Germany}

\date{1 September 2026}

\begin{abstract}
  Hyperdensity functional theory [Sammüller \textit{et al.}, \href{https://doi.org/10.1103/PhysRevLett.133.098201}{Phys.\ Rev.\ Lett.\ \textbf{133}, 098201 (2024)}] facilitates the investigation of the equilibrium behavior of a general order parameter or statistical mechanical observable in spatially inhomogeneous classical many-body systems.
  The approach is based on applying the exact Mermin-Evans classical density functional mapping to an extended ensemble.
  Here we present the multivariate generalization for investigating simultaneously the properties and interrelations of several different hyperobservables of choice.
  The resulting framework gives rise to a systematic characterization and prediction scheme for general many-body phenomena.
  All pertinent equilibrium averages, variances, and covariances constitute universal density functionals, as we demonstrate explicitly.
  Associated one-body hyperfluctuation profiles quantify the degree of correlation of the local density with first- and second-order combinations of hyperobservables.
  These multivariate hyperfluctuation profiles are accessible in many-body simulations and they satisfy exact hyper-Ornstein-Zernike equations, which we derive from the minimization principle in the extended multivariate ensemble.
  The formal structure of the theory integrates naturally with supervised machine learning, which renders all hyperdensity functionals accessible in practice via training of neural networks on simulation data.
  We demonstrate all salient techniques using the illustrative case of clustering in confined hard rod fluids, thereby choosing the total number of particles and the largest cluster size as the representative hyperobservables of interest.
  Our numerical methodology enables the efficient and successful prediction of all statistical quantities induced by the chosen hyperobservables, which we verify via comparison to test data and which we attribute to the tight interplay of first-principles and machine-learning concepts that our general approach combines.
\end{abstract}

\maketitle

\section{Introduction}
\label{sec:Introduction}

Classical density functional theory \cite{EvansNatureLiquidvapourInterface1979,EvansDensityFunctionalsTheory1992,EvansNewDevelopmentsClassical2016,HansenTheorySimpleLiquids2013,SchmidtPowerFunctionalTheory2022} is a powerful approach for treating the collective behavior of many-body systems.
The theory is based on using the one-body density distribution $\rho(\rv)$, where $\rv$ indicates position, as its central variable.
The formally exact Mermin-Evans minimization principle \cite{MerminThermalPropertiesInhomogeneous1965,EvansNatureLiquidvapourInterface1979} ascertains that the grand potential functional $\Omega[\rho]$ is minimized by the true equilibrium density distribution and that the value of the functional is the true equilibrium grand potential.
Going beyond the density ``profile'' $\rho(\rv)$, the local compressibility $\chi_\mu(\rv)=\partial \rho(\rv) / \partial \mu$, i.e., the partial derivative of the density profile with respect to the chemical potential $\mu$, was put forward \cite{EvansQuantifyingDensityFluctuations2015,EvansLocalCompressibilityLiquids2015,ChackoSolventFluctuationsAround2017,EvansUnifiedDescriptionHydrophilic2019,CoeDensityDepletionEnhanced2022,WildingWhatBestSimulation2024}, along with more general fluctuation profiles \cite{EckertFluctuationProfilesInhomogeneous2020,EckertLocalMeasuresFluctuations2023,CoeUnderstandingPhysicsHydrophobic2023}, in order to quantify and rationalize complex interfacial behavior, such as drying of a solvent near substrates and around solutes \cite{EvansQuantifyingDensityFluctuations2015,EvansLocalCompressibilityLiquids2015,ChackoSolventFluctuationsAround2017,EvansUnifiedDescriptionHydrophilic2019,CoeDensityDepletionEnhanced2022,CoeUnderstandingPhysicsHydrophobic2023,WildingWhatBestSimulation2024}.

Within classical density functional theory the complexity of the many-body problem is tucked away in the excess (over ideal gas) intrinsic free energy functional $F_\rmexc[\rho]$, which constitutes the nontrivial contribution to $\Omega[\rho]$ and which is unknown in general.
Recent progress on the basis of machine learning has propelled the decade-long quest for the construction of reliable approximations for $F_\rmexc[\rho]$ to a new and arguably very promising level.
Both classically \cite{Santos-SilvaNeuralnetworkApproachModeling2014,LinClassicalDensityFunctional2019,LinAnalyticalClassicalDensity2020,CatsMachinelearningFreeenergyFunctionals2021,QiaoEnhancingGasSolubility2020,YatsyshinPhysicsconstrainedBayesianInference2022,Malpica-MoralesPhysicsinformedBayesianInference2023,FangReliableEmulationComplex2022,DijkmanLearningNeuralFreeEnergy2025,KelleyBridgingElectronicClassical2024,BuiLearningClassicalDensity2025,SammullerNeuralFunctionalTheory2023,SammullerWhyNeuralFunctionals2024,SammullerNeuralDensityFunctionals2024,SammullerNeuralDensityFunctional2025,KampaMetadensityFunctionalTheory2025,RobitschkoLearningBulkInterfacial2025,SammullerDeterminingChemicalPotential2025} and within the quantum counterpart of the theory \cite{NagaiNeuralnetworkKohnShamExchangecorrelation2018,NagaiCompletingDensityFunctional2020,SchmidtMachineLearningPhysical2019,ZhouExactExchangeCorrelation2019,LiDeeplearningDensityFunctional2022,LiKohnShamEquationsRegularizer2021,GedeonMachineLearningDerivative2022,PedersonMachineLearningDensity2022}, as well as for nonequilibrium situations \cite{delasHerasPerspectiveHowOvercome2023,ZimmermannNeuralForceFunctional2024} based on power functional theory~\cite{SchmidtPowerFunctionalTheory2022}, a range of different methods has been put forward.
In particular, representing the one-body direct correlation functional \cite{HansenTheorySimpleLiquids2013,EvansNatureLiquidvapourInterface1979,EvansDensityFunctionalsTheory1992} locally by a standard feed-forward neural network \cite{SammullerNeuralFunctionalTheory2023,SammullerWhyNeuralFunctionals2024,SammullerNeuralDensityFunctionals2024,SammullerNeuralDensityFunctional2025,KampaMetadensityFunctionalTheory2025,RobitschkoLearningBulkInterfacial2025,SammullerDeterminingChemicalPotential2025,BuiLearningClassicalDensity2025} proves to be a versatile and computationally straightforward approach to use simulation results as training data for supervised machine learning of the underlying functional mapping; we refer the reader to Ref.~\cite{SammullerNeuralDensityFunctionals2024} for generalizations of this local learning technique and the relationship to other methods \cite{DijkmanLearningNeuralFreeEnergy2025,RamLearnedFreeenergyFunctionals2025}.

Tight integration of the fundamental statistical mechanics turns out to be a crucial aspect in many recent machine learning methods as well as in conceptual and numerical workflows.
Classical density functional theory in particular provides a rigorous mathematical footing that naturally suits the incorporation of neural networks and modern computational methods such as automatic differentiation.
Unlike in other common applications of machine learning techniques, the target mapping that is represented by a neural network does not constitute a quantity that is of immediate and direct interest in applications.
Instead, focusing on the arguably abstract yet fundamental one-body direct correlation functional allows one to keep the entire formal structure of classical density functional theory intact, thereby enabling a consistent and unified starting point for the determination of a wide range of desired quantities.
Exploiting underlying theoretical principles is hence crucial for conceiving accurate and physically interpretable machine learning methodologies.
Vice versa, the inclusion of machine learning triggers strong incentive to pursue fundamental theory development, as we demonstrate in the present work.

The recent hyperdensity functional theory \cite{SammullerHyperdensityFunctionalTheory2024,SammullerWhyHyperdensityFunctionals2025} constitutes a conceptual step beyond standard classical density functional theory \cite{EvansNatureLiquidvapourInterface1979,EvansDensityFunctionalsTheory1992,EvansNewDevelopmentsClassical2016,HansenTheorySimpleLiquids2013}, as it allows one to address the equilibrium average of a general, i.e., virtually arbitrary, observable of interest.
Such a ``hyperobservable'' is represented by a classical phase space function $\hat A$ and the theory facilitates to formulate its average as a universal density functional $A[\rho]$.
That the density profile alone is enough to determine the thermal mean of \emph{any} quantity $\hat A$ for a given fluid type, as specified by its intrinsic Hamiltonian and for prescribed temperature $T$, is an immediate mathematical consequence of the Mermin-Evans map \cite{MerminThermalPropertiesInhomogeneous1965,EvansNatureLiquidvapourInterface1979}.

Hyperdensity functional theory \cite{SammullerHyperdensityFunctionalTheory2024,SammullerWhyHyperdensityFunctionals2025} renders this merely formal theorem operational for the purpose of addressing concrete physical problems and phenomena.
This is achieved by formulating a route toward expressing $A[\rho]$ as a practically accessible density functional.
For a chosen hyperobservable $\hat A$, the theory gives rise to a corresponding hyperfluctuation profile $\chi_A(\rv)$, which correlates local density fluctuations with the instantaneous value of $\hat A$ for each microstate of the many-body phase space.
Furthermore, a hyperdirect correlation functional $c_A(\rv, [\rho])$ yields $A[\rho]$ upon functional line integration and this object is amenable to supervised machine learning.
An exact hyper-Ornstein-Zernike identity interrelates the one-body profiles $\rho(\rv)$, $\chi_A(\rv)$, and $c_A(\rv)$ in the form of a nontrivial integral equation.
In addition, hyperdensity functional theory is deeply connected with the hyperforce framework \cite{RobitschkoHyperforceBalanceThermal2024,MatthesGaugeInvarianceHyperforce2026,SammullerWhyHyperdensityFunctionals2025}, which arises as a consequence of gauge invariance \cite{MullerGaugeInvarianceEquilibrium2024,MullerWhyGaugeInvariance2025} and Noether's theorem in statistical mechanics \cite{HermannNoethersTheoremStatistical2021,HermannWhyNoethersTheorem2022,HermannVarianceFluctuationsNoether2022,HermannForceBalanceThermal2022,SammullerNoetherConstrainedCorrelationsEquilibrium2023,HermannNoetherInvarianceTheory2024,TschoppForceDensityFunctional2022,SammullerComparativeStudyForcebased2023}; see Ref.~\cite{RotenbergSymmetrySpottedStatistical2024} for a viewpoint and Ref.~\cite{MillerGaugeInvarianceApplies2025} for a popular account of these developments.

Here we investigate ramifications of the hyperdensity functional approach for the general multivariate case of a set of observables $\{\hat A_\ell\}$, where $\ell$ enumerates all items.
The multivariate extension is important as it allows one to study systematically the statistical interrelationships between several observables that are simultaneously of interest.
In particular, the generalization yields exhaustive insight into the characteristics of higher-order correlation functions and fluctuations, which is already pertinent for the case of a single observable and which constitutes a significant expansion of the concepts laid out in Refs.~\cite{SammullerHyperdensityFunctionalTheory2024,SammullerWhyHyperdensityFunctionals2025}.
Quantifying the emergent correlation structure forms a crucial prerequisite for identifying and rationalizing the causation of physical mechanisms in any considered many-body system.
Importantly, the multivariate hyperdensity functional formalism thereby rests firmly on first principles, which determines uniquely the specific forms and mutual relationships of the relevant correlation measures.

We exploit this formal structure of the theory to develop a corresponding multistage machine learning scheme for the iterative training of a hierarchy of neural hyperdirect correlation functionals.
These provide full quantitative closure of the theory, enabling fast and accurate predictions and opening up routes toward carrying out inverse design and high-throughput analysis tasks.
An indispensable conceptual and computational feature is the identification and utilization of G\^{a}teaux derivatives, which perform directional functional differentiation to capture the nonlocal correlations in key parts of the framework.
We leverage the immediate access to this important type of functional derivative via modern automatic differentiation techniques and develop efficient evaluation schemes in our numerical workflow.

The paper is organized as follows.
We present the formulation of the multivariate hyperdensity functional theory in Sec.~\ref{sec:Theory}, which encompasses a description of the extended statistical mechanical ensemble in Sec.~\ref{sec:ExtendendEnsemble}, the definition of hyperobservables and their associated hyperfluctuation profiles in Sec.~\ref{sec:Hyperobservables}, the derivation of first- and second-order hyper-Ornstein-Zernike equations and hyperdirect correlation functionals in Sec.~\ref{sec:HyperOZ}, a description of averages and covariances of hyperobservables expressed as density functionals in Sec.~\ref{sec:FunctionalIntegration}, and a practically relevant reformulation of the hyper-Ornstein-Zernike equations in terms of G\^{a}teaux derivatives in Sec.~\ref{sec:Gateaux}.
Consequences and applications of the theory are laid out in Sec.~\ref{sec:Applications}.
We discuss our multistage protocol for acquiring multivariate neural hyperdensity functionals via supervised machine learning in Sec.~\ref{sec:SimulationPerspective} and apply the framework for investigating the clustering statistics of the hard rod fluid in Sec.~\ref{sec:ClusterHardRods}.
We give conclusions and an outlook to future extensions and applications in Sec.~\ref{sec:Conclusions}.

\section{Theory}
\label{sec:Theory}

\subsection{Extended ensemble}
\label{sec:ExtendendEnsemble}

The system of interest is specified by its Hamiltonian $H$.
We consider an extended ensemble that is characterized by a set $\{\hat A_\ell\}$, where the index $\ell=1,\dots,L$ enumerates different observables $\hat A_\ell$, which may be virtually arbitrary phase space functions.
For terminological disambiguation to other (generic) observables, we will refer to the specific set $\{\hat A_\ell\}$ that pertains to the following derivations as ``hyperobservables''.
The extended grand ensemble is defined by the probability distribution function $f = \exp(-\beta(H-\mu N) + \sum_\ell \lambda_\ell\hat A_\ell))/\Xi$, where the normalization constant is the partition sum $\Xi=\Tr \exp(-\beta(H-\mu N) + \sum_\ell \lambda_\ell\hat A_\ell)$, with $N$ denoting the particle number and $\beta = 1 / (\kB T)$ with the Boltzmann constant $\kB$.
The operator $\Tr$ denotes the standard classical trace over phase space such that averages are defined as $\langle \cdot \rangle = \Tr \cdot\,f$.
The grand potential of the extended ensemble is $\Omega=-\kB T \ln\Xi$.

The equilibrium distribution function $f$ in the extended ensemble contains in its exponent a sum over all hyperobservables, with the coupling constants $\lambda_\ell$ determining the strength of the contribution of the $\ell$th hyperobservable.
However, we are interested ultimately only in the case of vanishing coupling, $\lambda_\ell = 0$ for all $\ell$, such that the equilibrium statistical physics that corresponds to the original Hamiltonian $H$ under consideration is restored.
Hence, when derivatives with respect to $\lambda_\ell$ are considered, it is generally implied that these are evaluated at $\lambda_\ell = 0$.
For higher-order derivatives, we take the corresponding convention that the evaluation at vanishing coupling occurs after the variation has been performed for all involved coupling parameters.

An analogous way to view the extended ensemble is to consider instead an extended Hamiltonian $H_A = H - \sum_\ell \lambda_\ell \hat A_\ell/\beta$ which enters the standard Boltzmann factor $\exp(-\beta(H_A-\mu N))$.
This setup leads to the identical partition sum and grand potential as described above.

\subsection{Hyperobservables and hyperfluctuation profiles}
\label{sec:Hyperobservables}

We wish to address the thermal expectation value $A_\ell$ of the $\ell$th hyperobservable as well as the covariance $A_{\ell\ells}$ of the $\ell$th and the $\ells$th hyperobservable, and hence define
\begin{align}
  \label{eq:AellDefinition}
  A_\ell &= \langle \hat{A}_\ell \rangle,\\
  \label{eq:AellellsDefinition}
  A_{\ell\ells} &= \cov(\hat{A}_\ell, \hat{A}_\ells),
\end{align}
where the covariance of two phase space functions $\hat X$ and $\hat Y$ is defined in the common way, $\cov(\hat X, \hat Y)=\langle \hat X \hat Y \rangle - XY$, with the averages $X= \langle \hat X \rangle$ and $Y= \langle \hat Y \rangle$.
Note that for $\ell = \ells$, Eq.~\eqref{eq:AellellsDefinition} reduces to $A_{\ell\ell} = \var(\hat{A}_\ell)$, where the variance of a phase space function $\hat X$ is defined as $\var(\hat X) = \langle \hat X^2 \rangle - X^2$.

Equivalently, $A_\ell$ and $A_{\ell\ells}$ can be expressed in terms of first and second derivatives of the grand potential with respect to the coupling parameters, which is straightforward to show via differentiating the elementary definition of $\Omega$:
\begin{align}
  \label{eq:AellAsParametricDerivative}
  A_\ell &= -\frac{\partial\beta\Omega}{\partial \lambda_\ell} \Big|_{V_\rmext},\\
  \label{eq:AellellsAsSimpleDerivative}
  A_{\ell\ells} &= -\frac{\partial \beta \Omega} {\partial\lambda_\ell \partial\lambda_\ells} \Big|_{V_\rmext}.
\end{align}
In the notation we have indicated that the form of the external potential $V_\rmext(\rv)$ is kept fixed upon building the parametric derivatives; also $\mu$ and $T$ are fixed (and suppressed in the notation).
By inserting Eq.~\eqref{eq:AellAsParametricDerivative} into Eq.~\eqref{eq:AellellsAsSimpleDerivative} one can straightforwardly show
\begin{equation}
  \label{eq:AellAsParametricDerivativeOfAells}
  A_{\ell\ells} = \frac{\partial A_\ell}{\partial \lambda_\ells} \Big|_{V_\rmext} = \frac{\partial A_\ells}{\partial \lambda_\ell} \Big|_{V_\rmext},
\end{equation}
which is reminiscent of a Maxwell identity.
The exchange symmetry of the indices $\ell$ and $\ells$ in Eq.~\eqref{eq:AellAsParametricDerivativeOfAells} is facilitated by exchanging the order of the two parametric derivatives acting on $\Omega$ in Eq.~\eqref{eq:AellellsAsSimpleDerivative}.

We define the following sets of first- and second-order hyperfluctuation profiles:
\begin{align}
  \label{eq:chiellAsCovariance}
  \chi_\ell(\rv) &= \cov(\hat{\rho}(\rv), \hat{A}_\ell),\\
  \label{eq:chiellellsAsTrivariance}
  \chi_{\ell\ells}(\rv) &= \triv(\hat{\rho}(\rv), \hat{A}_\ell, \hat{A}_\ells),
\end{align}
where $\hat{\rho}(\rv)=\sum_i \delta(\rv-\rv_i)$ is the standard one-body density ``operator'' with $\delta(\cdot)$ indicating the Dirac distribution and the sum running over all particles $i = 1,\dots,N$ that are currently in the system.
The first-order hyperfluctuation profile $\chi_\ell(\rv)$ is in direct generalization of the univariate case \cite{SammullerHyperdensityFunctionalTheory2024,SammullerWhyHyperdensityFunctionals2025}.
The second-order hyperfluctuation profile $\chi_{\ell\ells}(\rv)$ is given in Eq.~\eqref{eq:chiellellsAsTrivariance} in correlator form as the triple covariance (``trivariance''), which we define for three phase space functions $\hat X$, $\hat Y$, and $\hat Z$ as
\begin{equation}
  \label{eq:triv}
  \begin{split}
    \triv(\hat X, \hat Y, \hat Z) &= \langle \hat X \hat Y \hat Z \rangle- XYZ\\
                                  &\quad- \cov(\hat X, \hat Y)Z\\
                                  &\quad- \cov(\hat Y, \hat Z)X\\
                                  &\quad- \cov(\hat X, \hat Z)Y,
  \end{split}
\end{equation}
where $Z=\langle \hat Z \rangle$, which can alternatively be written as
\begin{equation}
  \label{eq:trivAlternative}
  \begin{split}
    \triv(\hat X, \hat Y, \hat Z) &= \langle \hat X \hat Y \hat Z \rangle + 2XYZ\\
                                  &\quad- \langle \hat X \hat Y \rangle Z - \langle \hat Y \hat Z \rangle X - \langle \hat X \hat Z \rangle Y.
  \end{split}
\end{equation}

More generally, the hyperfluctuation profiles can be defined as joint cumulants of $\hat{\rho}(\rv)$ with combinations of hyperobservables $\hat{A}_\ell, \hat{A}_\ells, \dots$, whereby the number of considered hyperobservables determines the order of the hyperfluctuation profile.
This definition is in line with the covariance and trivariance expressions~\eqref{eq:chiellAsCovariance} and \eqref{eq:chiellellsAsTrivariance}, and it allows one in principle to continue the hierarchy of hyperfluctuation profiles.
In the same fashion, the averages~\eqref{eq:AellDefinition} and covariances~\eqref{eq:AellellsDefinition} form the two lowest-order members of the hierarchy of joint cumulants of hyperobservables.
This general structure originates from the term $\exp(\sum_\ell \lambda_\ell \hat{A}_\ell)$ in the generalized Boltzmann factor, which renders the grand potential a multivariate cumulant generating function.
That the parametric derivative expressions \eqref{eq:AellAsParametricDerivative} and \eqref{eq:AellellsAsSimpleDerivative} hold confirms the fact that both $A_\ell$ and $A_{\ell\ells}$ constitute joint cumulants.

Similar considerations for the hyperfluctuation profiles give rise to
\begin{align}
  \label{eq:chiellAsDensityDerivative}
  \chi_\ell(\rv) &= \frac{\partial \rho(\rv)}{\partial \lambda_\ell} \Big|_{V_\rmext},\\
  \label{eq:chiellellsAsDensityDerivative}
  \chi_{\ell\ells}(\rv) &= \frac{\partial^2 \rho(\rv)}{\partial \lambda_\ell \partial \lambda_\ells} \Big|_{V_\rmext},
\end{align}
where the partial derivatives are performed again upon keeping $\mu, T$, and $V_\rmext(\rv)$ fixed.
Inserting Eq.~\eqref{eq:chiellAsDensityDerivative} into Eq.~\eqref{eq:chiellellsAsDensityDerivative} yields
\begin{equation}
  \chi_{\ell\ells}(\rv) = \frac{\partial \chi_\ell(\rv)}{\partial \lambda_\ells} \Big|_{V_\rmext} = \frac{\partial\chi_\ells(\rv)}{\partial\lambda_\ell} \Big|_{V_\rmext},
\end{equation}
whereby the labels $\ell$ and $\ells$ may again be exchanged due to rearranging the order of differentiation in Eq.~\eqref{eq:chiellellsAsDensityDerivative}, in analogy to Eq.~\eqref{eq:AellAsParametricDerivativeOfAells}.

The first-order hyperfluctuation profiles, as defined either in covariance form~\eqref{eq:chiellAsCovariance} or as parametric derivatives~\eqref{eq:chiellAsDensityDerivative}, are a generalization of the local compressibility~\cite{EvansLocalCompressibilityLiquids2015,EvansUnifiedDescriptionHydrophilic2019,EckertFluctuationProfilesInhomogeneous2020,EckertLocalMeasuresFluctuations2023,CoeUnderstandingPhysicsHydrophobic2023,WildingWhatBestSimulation2024,WildingCriticalSurfacePhase2025,WildingEmergentInterfacialLength2026} and the local thermal susceptibility \cite{EckertFluctuationProfilesInhomogeneous2020,EckertLocalMeasuresFluctuations2023,CoeUnderstandingPhysicsHydrophobic2023}.
These latter one-body profiles have been used extensively to address the physics of solvophobicity and hydrophobicity.
\citet{WildingCriticalSurfacePhase2025} have recently introduced the second-order pressure derivative of the density profile, $\partial^2 \rho(\rv) / \partial p^2$, for their investigation of surface criticality; our second-order hyperfluctuation profiles [Eqs.~\eqref{eq:chiellellsAsTrivariance} and \eqref{eq:chiellellsAsDensityDerivative}] can be seen as a formally analogous generalization.

As a precursor to the functional dependencies that we seek to identify, we express the first- and second-order hyperfluctuation profiles as the following functional derivatives:
\begin{align}
  \label{eq:chiellAsExternalPotentialDerivative}
  \chi_\ell(\rv) &= -\frac{\delta A_\ell}{\delta \beta V_\rmext(\rv)},\\
  \label{eq:chiellellsAsExternalPotentialDerivative}
  \chi_{\ell\ells}(\rv) &= -\frac{\delta A_{\ell\ells}}{\delta \beta V_\rmext(\rv)},
\end{align}
where we recall that $A_\ell$ is the average \eqref{eq:AellDefinition} of the hyperobservable $\hat A_\ell$ and $A_{\ell\ells}$ is the covariance \eqref{eq:AellellsDefinition} of $\hat{A}_\ell$ and $\hat{A}_\ells$.
Here $\delta/\delta \beta V_\rmext(\rv)$ denotes the functional derivative with respect to the scaled external potential $\beta V_\rmext(\rv)$; trivially one has $\delta/\delta\beta V_\rmext(\rv) = \beta^{-1}\delta/\delta V_\rmext(\rv)$ as $\beta = \mathrm{const}$.
Both Eqs.~\eqref{eq:chiellAsExternalPotentialDerivative} and \eqref{eq:chiellellsAsExternalPotentialDerivative} follow from functional differentiation of Eqs.~\eqref{eq:AellAsParametricDerivative} and \eqref{eq:AellellsAsSimpleDerivative} with respect to $-\beta V_\rmext(\rv)$, recalling $\rho(\rv) = \delta \Omega / \delta V_\rmext(\rv)$, exchanging the order of derivatives, and identifying the hyperfluctuation profiles via Eqs.~\eqref{eq:chiellAsDensityDerivative} and \eqref{eq:chiellellsAsDensityDerivative}.
Alternatively, one can start from the definition \eqref{eq:AellDefinition} of $A_\ell$, utilize the general identity $-\delta \langle \hat{X} \rangle/\delta \beta V_\rmext(\rv) = \cov(\hat{\rho}(\rv), \hat{X})$ for any phase space function $\hat X$, which can be verified by explicit calculation of the functional derivative \cite{EckertLocalMeasuresFluctuations2023}, and identify the covariance expression \eqref{eq:chiellAsCovariance} of $\chi_\ell(\rv)$.
Similarly, using the definition \eqref{eq:AellellsDefinition} of $A_{\ell\ells}$ and calculating the functional derivatives of the individual averages that contribute to the covariance then allows one to identify the trivariance expression \eqref{eq:chiellellsAsTrivariance} of $\chi_{\ell\ells}(\rv)$.

The thus developed relationships demonstrate the prowess of the extended ensemble to generate meaningful statistical mechanical quantities via parametric derivatives [Eqs.~\eqref{eq:AellAsParametricDerivative}, \eqref{eq:AellellsAsSimpleDerivative}, \eqref{eq:chiellAsDensityDerivative}, and \eqref{eq:chiellellsAsDensityDerivative}], which are associated to averages and covariances [Eqs.~\eqref{eq:AellDefinition} and \eqref{eq:AellellsDefinition}] as well as to functional derivatives [Eqs.~\eqref{eq:chiellAsExternalPotentialDerivative} and \eqref{eq:chiellellsAsExternalPotentialDerivative}] in the \emph{standard} grand ensemble.
All relationships are based on the elementary definition of the extended ensemble as described by its grand potential and we recall that we are ultimately interested only in the limit of vanishing coupling constants $\lambda_\ell$.

\subsection{Hyper-Ornstein-Zernike equations and hyperdirect correlation functionals}
\label{sec:HyperOZ}

We turn to the density functional description and assume that the extended Hamiltonian $H_A = H - \sum_\ell \lambda_\ell \hat{A}_\ell/\beta$ is sufficiently well-behaved for small values of $\lambda_\ell$ such that the corresponding grand potential density functional $\Omega[\rho]$ exists.
Then the Mermin-Evans minimization principle~\cite{MerminThermalPropertiesInhomogeneous1965,EvansNatureLiquidvapourInterface1979} of classical density functional theory ascertains that:
\begin{equation}
  \label{eq:minimizationPrinciple}
  \frac{\delta \Omega[\rho]}{\delta\rho(\rv)}\Big|_{H_A} = 0 \qquad \text{(min)},
\end{equation}
where $\delta/\delta\rho(\rv)$ denotes the functional derivative with respect to $\rho(\rv)$.
The extended Hamiltonian $H_A$, including the form of the external potential $V_\rmext(\rv)$, is kept fixed together with $\mu$ and $T$ upon building the density functional derivative.
Equality in the relationship \eqref{eq:minimizationPrinciple} is achieved for the true equilibrium density profile $\rho(\rv) = \langle \hat{\rho}(\rv) \rangle$.

The grand potential density functional $\Omega[\rho]$ splits into several distinct additive contributions according to
\begin{equation}
  \label{eq:omegaFunctionalSplitting}
  \Omega[\rho] = F_\rmid[\rho]+F_\rmexc[\rho]+\int d\rv \rho(\rv)[V_\rmext(\rv)-\mu],
\end{equation}
where the ideal gas free energy functional is given explicitly as $F_\rmid[\rho] = \kB T \int d\rv \rho(\rv) [\ln(\rho(\rv)\Lambda^d) - 1]$ with the thermal de Broglie wavelength $\Lambda$ and spatial dimensionality $d$; we set $\Lambda = 1$ in the following.
The intrinsic excess free energy functional $F_\rmexc[\rho]$ accounts for the nontrivial interparticle interactions and it is hence unknown in general.
Note that in our extended ensemble setting, the interaction effects can be interpreted to consist of both the unperturbed interparticle interaction potential $u(\rv^N)$ and the coupling to the extension, i.e., $u(\rv^N) - \sum_\ell \lambda_\ell \hat{A}_\ell/\beta$ in total.
Therefore, $F_\mathrm{exc}[\rho]$ depends implicitly on the coupling constants $\{\lambda_\ell\}$, which will be crucial in subsequent derivations.

Inserting the splitting \eqref{eq:omegaFunctionalSplitting} of $\Omega[\rho]$ into Eq.~\eqref{eq:minimizationPrinciple} leads to the standard Euler-Lagrange equation
\begin{equation}
  \label{eq:EulerLagrange}
  c_1(\rv,[\rho]) = \ln\rho(\rv) + \beta V_\rmext(\rv) - \beta \mu,
\end{equation}
where $c_1(\rv,[\rho]) = -\delta \beta F_\rmexc[\rho]/\delta\rho(\rv)$ is the one-body direct correlation functional \cite{HansenTheorySimpleLiquids2013,EvansNatureLiquidvapourInterface1979,EvansDensityFunctionalsTheory1992,EvansNewDevelopmentsClassical2016,SchmidtPowerFunctionalTheory2022}.
The algebraic form of Eq.~\eqref{eq:EulerLagrange} is thereby unchanged irrespective of working in the standard grand canonical or in the extended ensemble.

As the minimization principle \eqref{eq:minimizationPrinciple} holds true for any (permissible) form of $H_A$, we retain a valid identity by differentiating Eq.~\eqref{eq:minimizationPrinciple} with respect to $\lambda_\ell$ and hence
\begin{equation}
  \label{eq:minimizationPrincipleParametricDerivative}
  \frac{\partial}{\partial\lambda_\ell} \frac{\delta \Omega[\rho]}{\delta\rho(\rv)} \Big|_{H_A} = 0.
\end{equation}
Analogous to Eq.~\eqref{eq:minimizationPrincipleParametricDerivative} we can differentiate the Euler-Lagrange equation \eqref{eq:EulerLagrange} with respect to $\lambda_\ell$, thereby keeping $V_\rmext(\rv)$ and the thermodynamic conditions $\mu$, $T$ fixed, but monitoring the changes in the equilibrium density profile $\rho(\rv)$ upon changing $\lambda_\ell$; we recall the relationship \eqref{eq:chiellAsDensityDerivative} of $\chi_\ell(\rv)$ as is relevant for the use of the chain rule of differentiation.
The result after rearrangement is the following (first-order) hyper-Ornstein-Zernike relation \cite{SammullerHyperdensityFunctionalTheory2024,SammullerWhyHyperdensityFunctionals2025}:
\begin{equation}
  \label{eq:hyperOZ}
  c^A_{1,\ell}(\rv,[\rho]) = \frac{\chi_\ell(\rv)}{\rho(\rv)} - \int d\rv' \chi_\ell(\rv') c_2(\rv,\rv',[\rho]).
\end{equation}
The (first-order) one-body hyperdirect correlation functional $c_{1,\ell}^A(\rv,[\rho])$ is thereby given as:
\begin{equation}
  \label{eq:cA1ellAsParametricDerivative}
  c^A_{1,\ell}(\rv,[\rho]) = \frac{\partial c_1(\rv,[\rho])}{\partial \lambda_\ell}\Big|_\rho,\\
\end{equation}
where the parametric derivative is taken at fixed density profile as is indicated in the notation.
Note that the hyperdirect correlation functional $c^A_{1,\ell}(\rv,[\rho])$ only tracks the \emph{direct} effect of varying $\lambda_\ell$, whereas changes of the equilibrium density profile $\rho(\rv)$ that are induced by changing $\lambda_\ell$ are captured by the second term on the right-hand side of Eq.~\eqref{eq:hyperOZ}, which introduces coupling via the standard two-body direct correlation density functional $c_2(\rv,\rv',[\rho])=\delta c_1(\rv,[\rho])/\delta \rho(\rv')$.
Crucially, $c^A_{1,\ell}(\rv,[\rho])$ as defined via Eq.~\eqref{eq:cA1ellAsParametricDerivative} carries density functional dependence, as (i) $c_1(\rv,[\rho])$ is a universal density functional and (ii) the minimization principle continues to apply in the extended ensemble.
In Sec.~\ref{sec:Gateaux}, we demonstrate an alternative formulation of the hyper-Ornstein-Zernike equation~\eqref{eq:hyperOZ} based on directional functional (G\^{a}teaux) derivatives.

In order to address the coupling between different hyperobservables we go one order further and hence differentiate Eq.~\eqref{eq:minimizationPrincipleParametricDerivative} with respect to a second parameter $\lambda_\ells$ at fixed thermodynamic state point $\mu$, $T$ and external potential $V_\mathrm{ext}(\rv)$.
Building this second (mixed) parametric derivative of the Euler-Lagrange equation \eqref{eq:EulerLagrange} is analogous to building the first derivative of the first-order hyper-Ornstein-Zernike relation \eqref{eq:hyperOZ} with respect to $\lambda_\ells$.
The result is in both cases the following second-order hyper-Ornstein-Zernike relation:
\begin{equation}
  \label{eq:hyperOZsecondOrder}
  \begin{split}
    c^A_{1,\ell\ells}(\rv,[\rho]) &= \frac{\chi_{\ell\ells}(\rv)}{\rho(\rv)} -\frac{\chi_\ell(\rv)\chi_\ells(\rv)}{\rho(\rv)^2}\\
                                  &\quad-\int d\rv' \Big[\chi_\ells(\rv')c^A_{2,\ell}(\rv,\rv') +\chi_\ell(\rv')c^A_{2,\ells}(\rv,\rv')\Big]\\
                                  &\quad-\int d\rv' d\rv'' \chi_\ell(\rv') \chi_\ells(\rv'') c_3(\rv,\rv',\rv'')\\
                                  &\quad-\int d\rv' \chi_{\ell\ells}(\rv') c_2(\rv,\rv').
  \end{split}
\end{equation}
We have organized Eq.~\eqref{eq:hyperOZsecondOrder} such that the terms on the right-hand side express the second-order one-body hyperdirect correlation functional
\begin{equation}
  \label{eq:cA1ellells}
  \begin{split}
    c^A_{1,\ell\ells}(\rv,[\rho]) &= \frac{\partial^2 c_1(\rv,[\rho])}{\partial\lambda_\ell\partial\lambda_\ells} \Big|_\rho\\
    &= \frac{\partial c^A_{1,\ell}(\rv,[\rho])}{\partial \lambda_\ells} \Big|_\rho,
  \end{split}
\end{equation}
where the second equality follows from Eq.~\eqref{eq:cA1ellAsParametricDerivative}.

For brevity of notation we have dropped the density functional dependences of $c^A_{2,\ell}(\rv,\rv',[\rho])$, $c_2(\rv,\rv',[\rho])$, and $c_3(\rv,\rv',\rv'',[\rho])$ on the right-hand side of Eq.~\eqref{eq:hyperOZsecondOrder}, but emphasize that these are indeed all density functionals.
The first-order two-body hyperdirect correlation functional is thereby defined as:
\begin{equation}
  \label{eq:cA2ellFromc2}
  c^A_{2,\ell}(\rv,\rv',[\rho]) = \frac{\partial c_2(\rv,\rv',[\rho])}{\partial \lambda_\ell} \Big|_\rho,
\end{equation}
in analogy to the one-body version \eqref{eq:cA1ellAsParametricDerivative}.
We can readily obtain the alternative form
\begin{equation}
  \label{eq:cA2ellFromcA1ell}
  c^A_{2,\ell}(\rv,\rv',[\rho]) = \frac{\delta c^A_{1,\ell}(\rv,[\rho])}{\delta\rho(\rv')}
\end{equation}
by using the standard definition $c_2(\rv,\rv',[\rho])=\delta c_1(\rv,[\rho])/\delta\rho(\rv')$ in Eq.~\eqref{eq:cA2ellFromc2}, reordering the parametric and the functional derivatives and then identifying $c^A_{1,\ell}(\rv,[\rho])$ via Eq.~\eqref{eq:cA1ellAsParametricDerivative}.
All integral expressions in the second-order hyper-Ornstein-Zernike equation~\eqref{eq:hyperOZsecondOrder} can be treated as directional functional derivatives using the G\^{a}teaux calculus laid out in Sec.~\ref{sec:Gateaux}.

\subsection{Hyperobservables as density functionals}
\label{sec:FunctionalIntegration}

We aim to find density functional expressions for both the mean of the $\ell$th hyperobservable, $A_\ell$, and for the global covariance of the $\ell$th and the $\ells$th hyperobservable, $A_{\ell\ells}$.
We start with the relation \eqref{eq:AellAsParametricDerivative} of $A_\ell$ and express the parametric derivative on the right-hand side as the following parametric derivative of the excess free energy functional:
\begin{equation}
  \label{eq:derivationAcov1}
  \frac{\partial\Omega[\rho]}{\partial\lambda_\ell}\Big|_{V_\rmext} = \frac{\partial F_\rmexc[\rho]}{\partial\lambda_\ell} \Big|_{\rho},
\end{equation}
where the form of the density profile $\rho(\rv)$ is kept fixed upon building the derivative of $F_\mathrm{exc}[\rho]$.
Equation~\eqref{eq:derivationAcov1} follows from using the chain rule and the splitting \eqref{eq:omegaFunctionalSplitting} of $\Omega[\rho]$.
The arising terms may be simplified by identifying the minimization principle \eqref{eq:minimizationPrinciple} as well as eliminating contributions independent of $\lambda_\ell$, which leads to the compact and arguably surprising result \eqref{eq:derivationAcov1}.
We refer the reader to Refs.~\cite{SammullerHyperdensityFunctionalTheory2024,SammullerWhyHyperdensityFunctionals2025} for further details of this derivation.

Using relation~\eqref{eq:derivationAcov1} in Eq.~\eqref{eq:AellAsParametricDerivative} gives rise to the thermal average in density functional form:
\begin{equation}
  \label{eq:AellFromFexc}
  A_\ell[\rho] = -\frac{\partial \beta F_\rmexc[\rho]}{\partial \lambda_\ell} \Big|_{\rho}.
\end{equation}
Due to being generated via a parametric derivative of $F_\rmexc[\rho]$ for the extended Hamiltonian, the thermal average $A_\ell[\rho]$ inherits the property of being a universal density functional, as indicated in the notation.

In practice it is useful to substantiate Eq.~\eqref{eq:AellFromFexc} by expressing the excess free energy in terms of the functional line integral \cite{EvansNatureLiquidvapourInterface1979,EvansDensityFunctionalsTheory1992}
\begin{equation}
  \label{eq:FexcAsFunctionalIntegral}
  \begin{split}
    F_\rmexc[\rho] &= -\kB T \int \mathcal{D}[\rho] c_1(\rv,[\rho])\\
                   &= -\kB T \int d\rv\rho(\rv)\int_0^1da c_1(\rv,[a\rho]).
  \end{split}
\end{equation}
The second line in Eq.~\eqref{eq:FexcAsFunctionalIntegral} constitutes a linear parametrization of the general functional line integral $\int \mathcal{D}[\rho]$ whereby $a\rho(\rv)$, $0 \leq a \leq 1$, is a scaled version of the ``target'' density profile $\rho(\rv)$ that appears as the functional argument on the left hand side.
Inserting Eq.~\eqref{eq:FexcAsFunctionalIntegral} into Eq.~\eqref{eq:AellFromFexc}, then moving the parametric derivative with respect to $\lambda_\ell$ inside of the integral and using the definition \eqref{eq:cA1ellAsParametricDerivative} of the first-order one-body hyperdirect correlation functional $c^A_{1,\ell}(\rv,[\rho])$ yields:
\begin{equation}
  \label{eq:AellAsFunctionalIntegral}
  A_\ell[\rho] =\int d\rv\rho(\rv)\int_0^1 da c^A_{1,\ell}(\rv,[a\rho]).
\end{equation}
From either Eq.~\eqref{eq:AellFromFexc} or Eq.~\eqref{eq:AellAsFunctionalIntegral}, one can further conclude that:
\begin{equation}
  \label{eq:cA1ellFromAell}
  c^A_{1,\ell}(\rv,[\rho]) = \frac{\delta A_\ell[\rho]}{\delta \rho(\rv)},
\end{equation}
which reveals that the first-order one-body hyperdirect correlation functional is a measure of (direct) changes of the thermal mean $A_\ell$ upon local changes in the density profile $\rho(\rv)$.

Equation~\eqref{eq:AellAsFunctionalIntegral} is of a form which facilitates explicit access to $A_\ell[\rho]$ provided that $c^A_{1,\ell}(\rv,[\rho])$ is known.
As demonstrated in Refs.~\cite{SammullerHyperdensityFunctionalTheory2024,SammullerWhyHyperdensityFunctionals2025} for a single hyperobservable and in Sec.~\ref{sec:Applications} for the multivariate case, a practical route to obtain the required hyperdirect correlation functionals $c^A_{1,\ell}(\rv,[\rho])$ proceeds via simulation-based machine learning of neural functionals \cite{SammullerNeuralFunctionalTheory2023,SammullerWhyNeuralFunctionals2024}.

We next turn to the covariance $A_{\ell\ells}$ and recall its relationship \eqref{eq:AellellsAsSimpleDerivative} to the second mixed parametric derivative of the grand potential at fixed form of the external potential.
We apply this operation to the grand potential density functional \eqref{eq:omegaFunctionalSplitting}, which is analogous to building the derivative of Eq.~\eqref{eq:derivationAcov1} with respect to $\lambda_\ells$.
The result is:
\begin{equation}
  \label{eq:OmegaSecondLambdaDerivativeRaw}
  \begin{split}
    \frac{\partial^2 \Omega[\rho]}{\partial \lambda_\ell\partial\lambda_\ells} \Big|_{V_\rmext} &= \frac{\partial^2 F_\rmexc[\rho]}{\partial \lambda_\ell\partial \lambda_\ells}\Big|_\rho\\
                                                                                                &\quad + \int d\rv' \chi_\ells(\rv') \frac{\delta}{\delta\rho(\rv')} \frac{\partial F_\rmexc[\rho]}{\partial\lambda_\ell}\Big|_\rho\\
                                                                                                &\quad + \frac{\delta}{\delta\rho(\rv')} \int d\rv \chi_\ell(\rv) \frac{\delta\Omega[\rho]}{\delta\rho(\rv)} \Big|_{V_\rmext},
  \end{split}
\end{equation}
where we address each individual contribution in the following.

The third term on the right-hand side of Eq.~\eqref{eq:OmegaSecondLambdaDerivativeRaw} can be simplified by moving the density functional derivative inside of the integral.
Then applying the product rule of differentiation gives two terms, both of which vanish, one due to the minimization principle \eqref{eq:minimizationPrinciple} and one due to proto-form~\eqref{eq:minimizationPrincipleParametricDerivative} of the first-order hyper-Ornstein-Zernike equation.
The second term on the right-hand side of Eq.~\eqref{eq:OmegaSecondLambdaDerivativeRaw} can be rewritten upon reordering the derivatives and successively identifying $c_1(\rv,[\rho])$ in its standard form and $c^A_{1,\ell}(\rv,[\rho])$ via Eq.~\eqref{eq:cA1ellAsParametricDerivative}.
Recalling the second parametric derivative relationship \eqref{eq:AellellsAsSimpleDerivative}, we obtain the desired covariance as a density functional in the following form:
\begin{equation}
  \label{eq:AellellsAsFexcDerivative}
  A_{\ell\ells}[\rho] = -\frac{\partial^2 \beta F_\rmexc[\rho]} {\partial\lambda_\ell\partial\lambda_\ells} \Big|_\rho +\int d\rv \chi_\ells(\rv)c_{1,\ell}^A(\rv,[\rho]),
\end{equation}
which holds per construction also for interchanged labels $\ell$ and $\ells$ as a nontrivial symmetry of the second term on the right-hand side.

Alternatively, one can rewrite the second term on the right-hand side of Eq.~\eqref{eq:AellellsAsFexcDerivative} by multiplying the first-order hyper-Ornstein-Zernike relation \eqref{eq:hyperOZ} by $\chi_\ells(\rv)$, exchanging indices $\ell$ and $\ells$, and then integrating over $\rv'$.
Insertion into Eq.~\eqref{eq:AellellsAsFexcDerivative} gives the density functional form:
\begin{equation}
  \label{eq:AellellsAsFexcDerivative2}
  \begin{split}
  A_{\ell\ells}[\rho] &= -\frac{\partial^2 \beta F_\rmexc[\rho]} {\partial\lambda_\ell\partial\lambda_\ells} \Big|_\rho +\int d\rv \frac{\chi_\ell(\rv)\chi_\ells(\rv)}{\rho(\rv)}\\
                       &\quad -\int d\rv d\rv' \chi_\ell(\rv)\chi_\ells(\rv') c_2(\rv,\rv',[\rho]),
  \end{split}
\end{equation}
which is equivalent to Eq.~\eqref{eq:AellellsAsFexcDerivative}.
Again it is useful to express the present second parametric derivative of the excess free energy functional via functional line integration.
We hence write $F_\rmexc[\rho]$ as the standard functional line integral \eqref{eq:FexcAsFunctionalIntegral} over $c_1(\rv,[\rho])$, exchange the orders of derivatives and integrals and identify the second-order one-body hyperdirect correlation functional $c^A_{1,\ell\ells}(\rv,[\rho])$ according to Eq.~\eqref{eq:cA1ellells}.
The result is the following functional line integral:
\begin{equation}
  \label{eq:FexcEllEllsAsIntegral}
  -\frac{\partial^2 \beta F_\rmexc[\rho]}{\partial \lambda_\ell\lambda_\ells} = \int d\rv\rho(\rv)\int_0^1 da c^A_{1,\ell\ells}(\rv,[a\rho]),
\end{equation}
with the scaled density profile $a\rho(\rv)$, $0 \leq a \leq 1$, constituting the functional argument of the second-order hyperdirect correlation functional.

Insertion of Eq.~\eqref{eq:FexcEllEllsAsIntegral} into the covariance density functional \eqref{eq:AellellsAsFexcDerivative} yields:
\begin{equation}
  \label{eq:AellellsAsFunctionalIntegral}
  \begin{split}
    A_{\ell\ells}[\rho] &= \int d\rv\rho(\rv)\int_0^1da c^A_{1,\ell\ells}(\rv,[a\rho])\\
                         &\quad+ \int d\rv \chi_\ells(\rv) c^A_{1,\ell}(\rv,[\rho]),
  \end{split}
\end{equation}
where again the indices $\ell$ and $\ells$ in the second term can be interchanged with no harm.
From Eq.~\eqref{eq:AellellsAsFexcDerivative2}, we further obtain the equivalent form
\begin{equation}
  \begin{split}
  \label{eq:AellellsAsFunctionalIntegral2}
  A_{\ell\ells}[\rho] &= \int d\rv\rho(\rv)\int_0^1da c^A_{1,\ell\ells}(\rv,[a\rho])\\
                      &\quad+ \int d\rv \frac{\chi_\ell(\rv)\chi_\ells(\rv)}{\rho(\rv)}\\
                      &\quad- \int d\rv d\rv' \chi_\ell(\rv)\chi_\ells(\rv') c_2(\rv,\rv',[\rho]).
  \end{split}
\end{equation}
In Appendix~\ref{appendix:cov}, we verify numerically the equivalence of Eq.~\eqref{eq:AellellsAsFunctionalIntegral} in its original form and with exchanged indices $\ell$ and $\ells$ as well as Eq.~\eqref{eq:AellellsAsFunctionalIntegral2}, using the neural functional workflow laid out in Sec.~\ref{sec:Applications}.
Equations~\eqref{eq:AellellsAsFexcDerivative} and \eqref{eq:AellellsAsFunctionalIntegral} further reveal that
\begin{equation}
  \label{eq:cA1ellellsFromAellells}
  c^A_{1,\ell\ells}(\rv,[\rho]) = \frac{\delta A_{\ell\ells}[\rho]}{\delta \rho(\rv)} - \int d\rv \chi_\ells(\rv) c^A_{1,\ell}(\rv,[\rho]).
\end{equation}
We note the appearance of the second term on the right-hand side due to the chain rule, which might not be gleaned at first sight from the first-order result~\eqref{eq:cA1ellFromAell}.

As an alternative derivation, the form~\eqref{eq:AellellsAsFunctionalIntegral} of the covariance density functional can be obtained from the relationship $A_{\ell\ells}=\partial A_\ell/\partial\lambda_\ells$, see Eq.~\eqref{eq:AellAsParametricDerivativeOfAells}, applied to the density functional $A_\ell[\rho]$ as given in Eq.~\eqref{eq:AellAsFunctionalIntegral}.
One hence needs to differentiate $A_\ell[\rho]$ with respect to $\lambda_\ells$ upon fixing $\mu$, $T$, and $V_\rmext(\rv)$.
Differentiating the parametric dependence of $c_{1,\ell}^A(\rv,[a\rho])$ on the right-hand side of Eq.~\eqref{eq:AellAsFunctionalIntegral} with respect to the coupling parameter generates the functional line integral over $c_{1,\ell\ells}^A(\rv,[\rho])$ in Eq.~\eqref{eq:AellellsAsFunctionalIntegral}.
The second term stems from the changes of the target density profile that forms the upper bound of the functional line integral in Eq.~\eqref{eq:AellAsFunctionalIntegral}.
Performing the derivative with respect to $\lambda_\ells$ cancels the functional line integration, evaluates its integrand at the target density, and gives according to Eq.~\eqref{eq:chiellAsDensityDerivative} a factor $\chi_\ells(\rv)$ from the chain rule.
The result is the second term in Eq.~\eqref{eq:AellellsAsFunctionalIntegral}, which we reemphasize is no longer a functional line integral:
While $c^A_{1,\ell}(\rv,[\rho])$ is indeed a density functional, no integration over the functional argument takes place.
Instead, the hyperdirect correlation functional is merely evaluated at density $\rho(\rv)$, and the result enters the standard position integral together with the hyperfluctuation profile $\chi_\ells(\rv)$.
We show in Sec.~\ref{sec:Applications} that the density functional $A_{\ell\ells}[\rho]$ is, analogous to $A_\ell[\rho]$, explicitly accessible via a multistage supervised machine learning workflow based on simulation data of a considered fluid in inhomogeneous environments.

The above process of expressing parametric derivatives of $\Omega[\rho]$ with respect to (multiple) coupling parameters $\lambda_\ell, \lambda_\ells, \dots$ in terms of density functionals can be applied to generate a hierarchy of increasing complexity.
As a representative example, differentiating Eq.~\eqref{eq:AellellsAsFunctionalIntegral} by $\lambda_\ellss$ leads to
\begin{equation}
  \begin{split}
    &A_{\ell\ells\ellss}[\rho] =\\
    &\int d\rv \rho(\rv) \int_0^1 da c_{1,\ell\ells\ellss}^A(\rv,[a\rho])\\
    &\quad+ \int d\rv \chi_\ellss(\rv) c^A_{1,\ell\ells}(\rv,[\rho])\\
    &\quad+ \int d\rv \big[\chi_{\ells\ellss}(\rv) c_{1,\ell}^A(\rv,[\rho]) + \chi_\ells(\rv) c_{1,\ell\ellss}^A(\rv,[\rho]) \big]\\
    &\quad+ \int d\rv d\rv' \chi_\ells(\rv) \chi_\ellss(\rv') c_{2,\ell}^A(\rv,\rv',[\rho]).
  \end{split}
\end{equation}
where $A_{\ell\ells\ellss}=\triv(\hat A_\ell, \hat A_\ells, \hat A_\ellss)=\partial A_{\ell\ells}/\partial \lambda_\ellss$ and $c^A_{1,\ell\ells\ellss}(\rv,[\rho])=\partial c^A_{1,\ell\ells}(\rv,[\rho])/\partial \lambda_\ellss$.

For completeness, the case of a single hyperobservable $\hat A = \hat A_1$ of interest is contained in the current theory simply via setting the number of hyperobservables $L=1$.
Then the sole remaining variance is $A_{11}= \var(\hat A)$.
The second-order hyperfluctuation profile acquires corresponding meaning as $\chi_{11}(\rv)=\triv(\hat\rho(\rv),\hat A, \hat A) = \cov(\hat\rho(\rv),\hat A^2)-2\chi_1(\rv)A$.
The second-order hyper-Ornstein-Zernike equation~\eqref{eq:hyperOZsecondOrder} can then be simplified as follows:
\begin{equation}
  \begin{split}
  c^A_{1,11}(\rv,[\rho]) &= \frac{\chi_{11}(\rv)}{\rho(\rv)} -\frac{\chi_1(\rv)^2}{\rho(\rv)^2}\\
                     &\quad- 2\int d\rv' c^A_{2,1}(\rv,\rv',[\rho])\chi_1(\rv'),\\
                     &\quad- \int d\rv'd\rv'' c_3(\rv,\rv',\rv'',[\rho])\chi_1(\rv')\chi_1(\rv'')\\
                     &\quad- \int d\rv' c_2(\rv,\rv',[\rho])\chi_{11}(\rv'),
  \end{split}
\end{equation}
which extends the univariate theory of Refs.~\cite{SammullerHyperdensityFunctionalTheory2024,SammullerWhyHyperdensityFunctionals2025} to the second order.

\subsection{Hyper-Ornstein-Zernike equations and G\^{a}teaux derivatives}
\label{sec:Gateaux}

While the derivations of the previous sections give a complete account of the multivariate hyperdensity functional theory, we reexamine in the following a formalization of the functional calculus, which enables a significant reduction in complexity for both theoretical and practical considerations.
An equivalent reformulation of integral expressions in the hyper-Ornstein-Zernike equations \eqref{eq:hyperOZ} and \eqref{eq:hyperOZsecondOrder} in terms of generalized directional functional derivatives is laid out here.
This sheds further light onto their underlying algebraic structure and proves to be crucial in subsequent numerical applications.

Starting with the first-order hyper-Ornstein-Zernike equation~\eqref{eq:hyperOZ}, we recall that the integral term on its right-hand side arises from the derivation via Eq.~\eqref{eq:minimizationPrincipleParametricDerivative} due to the chain rule of functional differentiation.
Fixing the free position argument $\rv$, the general structure of this prototypical expression is
\begin{equation}
  \label{eq:FunctionalDifferentialAsIntegral}
  \mathcal{D}_\psi \mathcal{F}[\rho] = \int d\rv' \psi(\rv') \frac{\delta \mathcal{F}[\rho]}{\delta \rho(\rv')}
\end{equation}
with generic function $\psi(\rv')$ and generic functional $\mathcal{F}[\rho]$ (not to be confused with the intrinsic free energy functional $F[\rho]$).
The second term in Eq.~\eqref{eq:hyperOZ} is recovered explicitly by setting $\psi(\rv') = \chi_\ell(\rv')$ and $\mathcal{F}[\rho] = c_1(\rv, [\rho])$ in Eq.~\eqref{eq:FunctionalDifferentialAsIntegral}.

By definition of the functional derivative \cite{ParrDensityFunctionalTheory1994}, Eq.~\eqref{eq:FunctionalDifferentialAsIntegral} may equivalently be expressed as the following G\^{a}teaux differential:
\begin{equation}
  \label{eq:FunctionalDifferentialAsGateaux}
  \mathcal{D}_\psi \mathcal{F}[\rho] = \frac{d \mathcal{F}[\rho + \epsilon \psi]}{d \epsilon} \Big|_{\epsilon = 0},
\end{equation}
where $\epsilon$ is a scalar parameter.
Equation~\eqref{eq:FunctionalDifferentialAsGateaux} makes apparent that terms of the form \eqref{eq:FunctionalDifferentialAsIntegral} constitute (generalized) directional derivatives of the functional $\mathcal{F}$ with respect to variations along $\psi$, evaluated at the functional argument $\rho$.
The corresponding directional (G\^{a}teaux) derivative operator $\mathcal{D}_\psi$ is introduced for succinct notation.

Evaluating terms of the form of the right-hand side of Eq.~\eqref{eq:FunctionalDifferentialAsIntegral} via directional differentiation according to Eq.~\eqref{eq:FunctionalDifferentialAsGateaux} is both straightforward and highly beneficial from a practical standpoint in numerical work.
Automatic differentiation techniques facilitate immediate implementation of Eq.~\eqref{eq:FunctionalDifferentialAsGateaux}; we give an example in the programming language Julia in Appendix~\ref{appendix:GateauxJulia}.
Utilizing this approach circumvents entirely the need to evaluate both the full functional derivative $\delta \mathcal{F}[\rho] / \delta \rho(\rv')$ as well as the spatial integral in Eq.~\eqref{eq:FunctionalDifferentialAsIntegral}.
Instead, only the variation with respect to a given perturbation $\epsilon \psi(\rv')$ must be evaluated in Eq.~\eqref{eq:FunctionalDifferentialAsGateaux}, which leads to a substantial reduction in both computational and memory demand.
As we demonstrate below, besides the strikingly compact code, the benefit lies in excellent performance.
The approach is similar to calculating Jacobian-vector products directly, without intermediate evaluation of the entire Jacobian, as is key strategy in various numerical routines that arise particularly in machine learning disciplines \cite{BaydinAutomaticDifferentiationMachine2018}.

In the second-order hyper-Ornstein-Zernike equation~\eqref{eq:hyperOZsecondOrder}, the integral term featuring $c_3(\rv, \rv', \rv'', [\rho])$ is of type
\begin{equation}
  \label{eq:FunctionalDifferentialAsIntegral2}
   \mathcal{D}_{\psi_1}\mathcal{D}_{\psi_2}\mathcal{F}[\rho] = \int d\rv' \int d\rv'' \psi_1(\rv') \psi_2(\rv'') \frac{\delta^2 \mathcal{F}[\rho]}{\delta \rho(\rv') \delta \rho(\rv'')},
\end{equation}
which corresponds to a second-order G\^{a}teaux derivative that is readily computed by a two-step application of Eq.~\eqref{eq:FunctionalDifferentialAsGateaux}.
A code example which facilitates the efficient evaluation of such generalized vector-Hessian-vector products is given in Appendix~\ref{appendix:GateauxJulia}.

For completeness, we spell out the reformulated versions of the first- and second-order hyper-Ornstein-Zernike equations \eqref{eq:hyperOZ} and \eqref{eq:hyperOZsecondOrder}, where the arising integrals are reexpressed in terms of G\^{a}teaux differentials:
\begin{align}
  \label{eq:hyperOZGateaux}
  c^A_{1,\ell}(\rv,[\rho]) &= \frac{\chi_\ell(\rv)}{\rho(\rv)} - \mathcal{D}_{\chi_\ell} c_1(\rv,[\rho]),\\
  \label{eq:hyperOZsecondOrderGateaux}
  \begin{split}
    c^A_{1,\ell\ells}(\rv,[\rho]) &= \frac{\chi_{\ell\ells}(\rv)}{\rho(\rv)} -\frac{\chi_\ell(\rv)\chi_\ells(\rv)}{\rho(\rv)^2}\\
                                  &\quad-\mathcal{D}_{\chi_\ells} c^A_{1,\ell}(\rv, [\rho]) - \mathcal{D}_{\chi_\ell} c^A_{1,\ells}(\rv, [\rho])\\
                                  &\quad-\mathcal{D}_{\chi_\ell} \mathcal{D}_{\chi_\ells} c_1(\rv, [\rho]) - \mathcal{D}_{\chi_{\ell\ells}} c_1(\rv, [\rho]).
  \end{split}
\end{align}
We demonstrate below that Eqs.~\eqref{eq:hyperOZGateaux} and \eqref{eq:hyperOZsecondOrderGateaux} constitute not merely formal rewritings, but rather crucial tools for dealing with the hyper-Ornstein-Zernike structure in numerical work.

\section{Applications}
\label{sec:Applications}

\subsection{Simulation and machine learning perspective}
\label{sec:SimulationPerspective}

\begin{figure*}
  \includegraphics{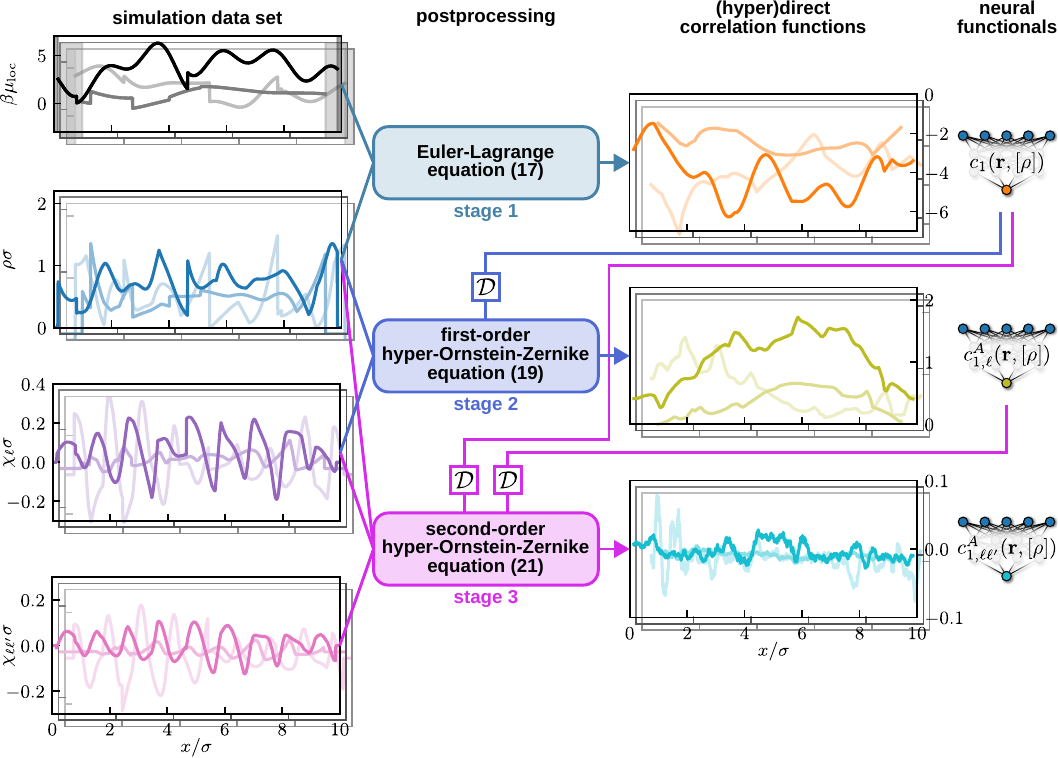}
  \caption{
    Multistage workflow for machine learning multivariate hyperdensity functionals.
    A grand canonical Monte Carlo simulation data set (left column) of equilibria of a considered fluid type serves as ground truth.
    Inhomogeneity is induced in each simulation by a randomly chosen (scaled) local chemical potential $\beta \mu_\mathrm{loc}(\rv) = \beta \mu - \beta V_\mathrm{ext}(\rv)$ and the resulting set of one-body profiles $\rho(\rv)$, $\chi_\ell(\rv)$, and $\chi_{\ell\ells}(\rv)$ is sampled.
    We depict exemplary data for the one-dimensional hard rod fluid where $\rv = x$.
    As hyperobservables of choice, we select the total number of particles, $\hat A_1$, and the number of particles in the largest cluster, $\hat A_2$, see the text for a detailed specification.
    Postprocessing of the simulation data occurs in three subsequent stages via the Euler-Lagrange equation~\eqref{eq:EulerLagrange}, the first-order hyper-Ornstein-Zernike equation~\eqref{eq:hyperOZ}, and the second-order hyper-Ornstein-Zernike equation~\eqref{eq:hyperOZsecondOrder}.
    The latter two equations are thereby implemented via Eqs.~\eqref{eq:hyperOZGateaux} and \eqref{eq:hyperOZsecondOrderGateaux} with the symbol $\mathcal{D}$ denoting the evaluation of directional functional (G\^{a}teaux) derivatives.
    The data set features hyperfluctuation and hyperdirect correlation profiles for all pertinent index combinations; we only show representative results for $\ell = 2$ and $\ells = 1$.
    The resulting (hyper)direct correlation profiles are the target quantities for machine learning corresponding neural (hyper)density functionals $c_1(\rv, [\rho])$, $c^A_\ell(\rv, [\rho])$, and $c^A_{\ell\ells}(\rv, [\rho])$, which are fit for application in both standard and hyperdensity functional theory.
    We show the workflow for neural functional predictions in Fig.~\ref{fig:prediction_example} as well as benchmark results in Fig.~\ref{fig:predictions}.
  }
  \label{fig:overview}
\end{figure*}

In the following, we lay out our multistage approach for machine learning the direct and hyperdirect correlation functionals based on simulation data for inhomogeneous fluid equilibria.
In Fig.~\ref{fig:overview} the general workflow is illustrated for a concrete application scenario, see Sec.~\ref{sec:ClusterHardRods}.
The neural functionals provide full access to averages $A_\ell[\rho]$ and covariances $A_{\ell\ells}[\rho]$ as well as to spatially resolved first- and second-order hyperfluctuation profiles $\chi_\ell(\rv)$ and $\chi_{\ell\ells}(\rv)$, as demonstrated below.

Let us first consider a specific inhomogeneous system of interest at fixed thermodynamic state point $\mu$, $T$ and with prescribed external potential $V_\mathrm{ext}(\rv)$ to reexamine the accessibility of the different terms that arise in the hyper-Ornstein-Zernike relations \eqref{eq:hyperOZ} and \eqref{eq:hyperOZsecondOrder}, cf.\ also their G\^{a}teaux forms~\eqref{eq:hyperOZGateaux} and \eqref{eq:hyperOZsecondOrderGateaux}.
For capturing the right-hand side of the first-order hyper-Ornstein-Zernike relation \eqref{eq:hyperOZ}, one requires to have results for the density profile $\rho(\rv)$ and the hyperfluctuation profiles $\chi_\ell(\rv)$ that arise in this system.
The former quantity is straightforward to acquire in standard grand canonical Monte Carlo simulations and the latter is available via sampling the covariance \eqref{eq:chiellAsCovariance} and crucially without having to represent the extended ensemble directly.
From these results, the first-order one-body hyperdirect correlation function $c^A_{1,\ell}(\rv)$ that pertains to the specific system under investigation can be constructed explicitly via the first-order hyper-Ornstein-Zernike equation.
For this, one either evaluates Eq.~\eqref{eq:hyperOZ}, where the standard two-body direct correlation functional $c_2(\rv,\rv',[\rho])$ can be obtained via (automatic) differentiation of a neural (or, if available, analytical) one-body direct correlation functional \cite{SammullerNeuralFunctionalTheory2023,SammullerWhyNeuralFunctionals2024}.
Alternatively and more efficiently, Eq.~\eqref{eq:hyperOZGateaux} facilitates an equivalent calculation using automatic directional derivatives of $c_1(\rv,[\rho])$.

For each system in the simulation data set, the measured quantities $\rho(\rv)$ and $\chi_\ell(\rv)$ as well as the density functional $c_1(\rv,[\rho])$ are hence required in order to evaluate the corresponding profiles $c^A_{1,\ell}(\rv)$.
Construction of a neural functional via training on the thus acquired data set is feasible to obtain an accurate representation of the universal density functional mapping $c^A_{1,\ell}(\rv,[\rho])$.
Expressing the average $A_\ell[\rho]$ as a density functional is then viable by employing the neural hyperdirect correlation functional in the functional line integral \eqref{eq:AellAsFunctionalIntegral}.
In Refs.~\cite{SammullerHyperdensityFunctionalTheory2024,SammullerWhyHyperdensityFunctionals2025}, applications of the first-order theory are laid out for a single hyperobservable, confirming the practical accessibility of all quantities as well as the possibility to machine learn the underlying functional mapping.

We turn to the second-order theory and consider the right-hand side of the second-order hyper-Ornstein-Zernike relation, cf.\ Eqs.~\eqref{eq:hyperOZsecondOrder} and \eqref{eq:hyperOZsecondOrderGateaux}, where the density profile $\rho(\rv)$ and the hyperfluctuation profiles $\chi_\ell(\rv)$ and $\chi_\ells(\rv)$ are available in analogy to the first-order case.
The second-order hyperfluctuation profile $\chi_{\ell\ells}(\rv)$ can be obtained via the trivariance correlator \eqref{eq:chiellellsAsTrivariance}, which must hence be sampled in all simulations.
While the trivariance does converge more slowly than covariances or standard averages, we emphasize that for scalar observables $\hat A_\ell$ and $\hat A_\ells$ of interest, $\chi_{\ell\ells}(\rv)$ is still a one-body profile depending only on a single position argument $\rv$.
This is beneficial from a practical point of view as it keeps the sampling feasible with standard histogram methods.

While both Eqs.~\eqref{eq:hyperOZsecondOrder} and \eqref{eq:hyperOZsecondOrderGateaux} are equivalent formulations of the second-order hyper-Ornstein-Zernike relation, their numerical treatment differs drastically such that only the latter Eq.~\eqref{eq:hyperOZsecondOrderGateaux} constitutes the computationally feasible route, as we illustrate in the following.
Dealing with Eq.~\eqref{eq:hyperOZsecondOrder} would formally require to evaluate all integral expressions on its right-hand side, thereby utilizing $c_2(\rv, \rv', [\rho])$, $c_3(\rv, \rv', \rv'', [\rho])$, and $c^A_{2,\ell}(\rv, \rv', [\rho])$ as generalized integral kernels.
The two- and three-body direct correlation functionals $c_2(\rv, \rv', [\rho])$ and $c_3(\rv, \rv', \rv'', [\rho])$ can be obtained in principle from automatic differentiation of $c_1(\rv, [\rho])$ \cite{SammullerNeuralFunctionalTheory2023,SammullerWhyNeuralFunctionals2024,SammullerNeuralDensityFunctional2025}, and $c^A_{2,\ell}(\rv, \rv', [\rho])$ follows from automatic differentiation of $c^A_{1,\ell}(\rv, [\rho])$ according to Eq.~\eqref{eq:cA2ellFromcA1ell}.
In (quasi-)one-dimensional geometries and on the first-order level of the theory, the evaluation of Eq.~\eqref{eq:hyperOZ} remains feasible via this direct route, as demonstrated in Refs.~\cite{SammullerHyperdensityFunctionalTheory2024,SammullerWhyHyperdensityFunctionals2025}.
However, if spatial resolution is increased, e.g.\ when working in multiscale settings \cite{SammullerNeuralFunctionalTheory2023} or in more general geometries \cite{GlitschNeuralDensityFunctional2025,WeimarCubicEquivariantNeural2026,ChengEquivariantLearningTransferable2026,KampaSphericalMetadensityFunctional2026}, the explicit evaluation of the full objects $c_2(\rv, \rv', [\rho])$, $c_3(\rv, \rv', \rv'', [\rho])$, and $c^A_{2,\ell}(\rv, \rv', [\rho])$ becomes increasingly expensive.
Concretely, for a discretized spatial grid with $n$ grid points, $c_2(\rv, \rv', [\rho])$ comprises $n^2$ values which need to be computed and stored in a naive approach.
On the second-order level of the theory, this issue becomes prohibitively expensive.
For $c_3(\rv, \rv', \rv'', [\rho])$, which appears in Eq.~\eqref{eq:hyperOZsecondOrder}, $n^3$ entries would need to be evaluated and stored, thereby using computationally demanding second-order (e.g.\ forward-over-reverse) automatic differentiation techniques \cite{BaydinAutomaticDifferentiationMachine2018}.
To circumvent this issue, we utilize the reformulation of the integral expressions in the hyper-Ornstein-Zernike equations~\eqref{eq:hyperOZ} and \eqref{eq:hyperOZsecondOrder} in terms of directional functional derivatives, see Eqs.~\eqref{eq:hyperOZGateaux} and \eqref{eq:hyperOZsecondOrderGateaux}.
Exploiting the G\^{a}teaux calculus enables a numerically efficient route toward evaluating all arising terms in both hyper-Ornstein-Zernike equations; we refer to Appendix~\ref{appendix:GateauxJulia} for implementation details.

Analogous to the first-order case, the resulting data set for $c^A_{1,\ell\ells}(\rv)$ evaluated with the second-order hyper-Ornstein-Zernike equation across all simulated systems provides the basis for extracting the underlying density functional relationship $c^A_{1,\ell\ells}(\rv,[\rho])$ via a neural network.
The covariances $A_{\ell\ells}[\rho]$ are then accessible as genuine density functionals by evaluating either Eq.~\eqref{eq:AellellsAsFunctionalIntegral} or Eq.~\eqref{eq:AellellsAsFunctionalIntegral2}, whereby the second-order one-body hyperdirect correlation functional enters in the functional line integration, and the remaining functionals are available from either first-order hyperdensity [$c^A_{1,\ell}(\rv,[\rho])$] or standard density functional theory [$c_2(\rv,\rv',[\rho])$].
Similarly, the first-order hyperfluctuation profiles $\chi_\ell(\rv)$ that appear in Eq.~\eqref{eq:hyperOZsecondOrder} may be obtained by solution of Eq.~\eqref{eq:hyperOZ}, see also the account given in Refs.~\cite{SammullerHyperdensityFunctionalTheory2024,SammullerWhyHyperdensityFunctionals2025}.
We recall the illustration of the multistage machine learning of neural hyperdensity functionals in Fig.~\ref{fig:overview}, and exemplify the workflow for the investigation of clustering statistics in the hard rod fluid, as laid out in Sec.~\ref{sec:ClusterHardRods}.

From the standpoint of having all occurring density functionals available, either as analytic approximation or via representation in terms of neural networks, both hyper-Ornstein-Zernike relations~\eqref{eq:hyperOZ} and \eqref{eq:hyperOZsecondOrder} form defining integral equations for the first- and second-order hyperfluctuation profiles.
Specifically, for a system with known density profile, as follows from solution of the Euler-Lagrange equation~\eqref{eq:EulerLagrange}, the first-order hyper-Ornstein-Zernike relation \eqref{eq:hyperOZ} is an integral equation for $\chi_\ell(\rv)$, in which the specific form of $c^A_{1,\ell}(\rv,[\rho])$, evaluated at the equilibrium density profile $\rho(\rv)$, constitutes a fixed inhomogeneity.
Having all $\chi_\ell(\rv)$ obtained in this way, the second-order hyper-Ornstein-Zernike relation \eqref{eq:hyperOZsecondOrder} then forms an integral equation for the second-order hyperfluctuation profiles $\chi_{\ell\ells}(\rv)$.
These appear both locally [first term on the right-hand side of Eq.~\eqref{eq:hyperOZsecondOrder}] and inside of the integration over $c_2(\rv,\rv',[\rho])$ [last term on the right-hand side of Eq.~\eqref{eq:hyperOZsecondOrder}].
All other terms are fixed and constitute the inhomogeneity that ultimately generates the resulting equilibrium shape of $\chi_{\ell\ells}(\rv)$.
The prediction workflow from neural hyperdensity functionals to averages, covariances, and hyperfluctuation profiles is illustrated in Fig.~\ref{fig:prediction_example}.

\subsection{Case study: Fluctuations of particle number and cluster size in confined hard rod fluids}
\label{sec:ClusterHardRods}

\begin{figure*}
  \includegraphics{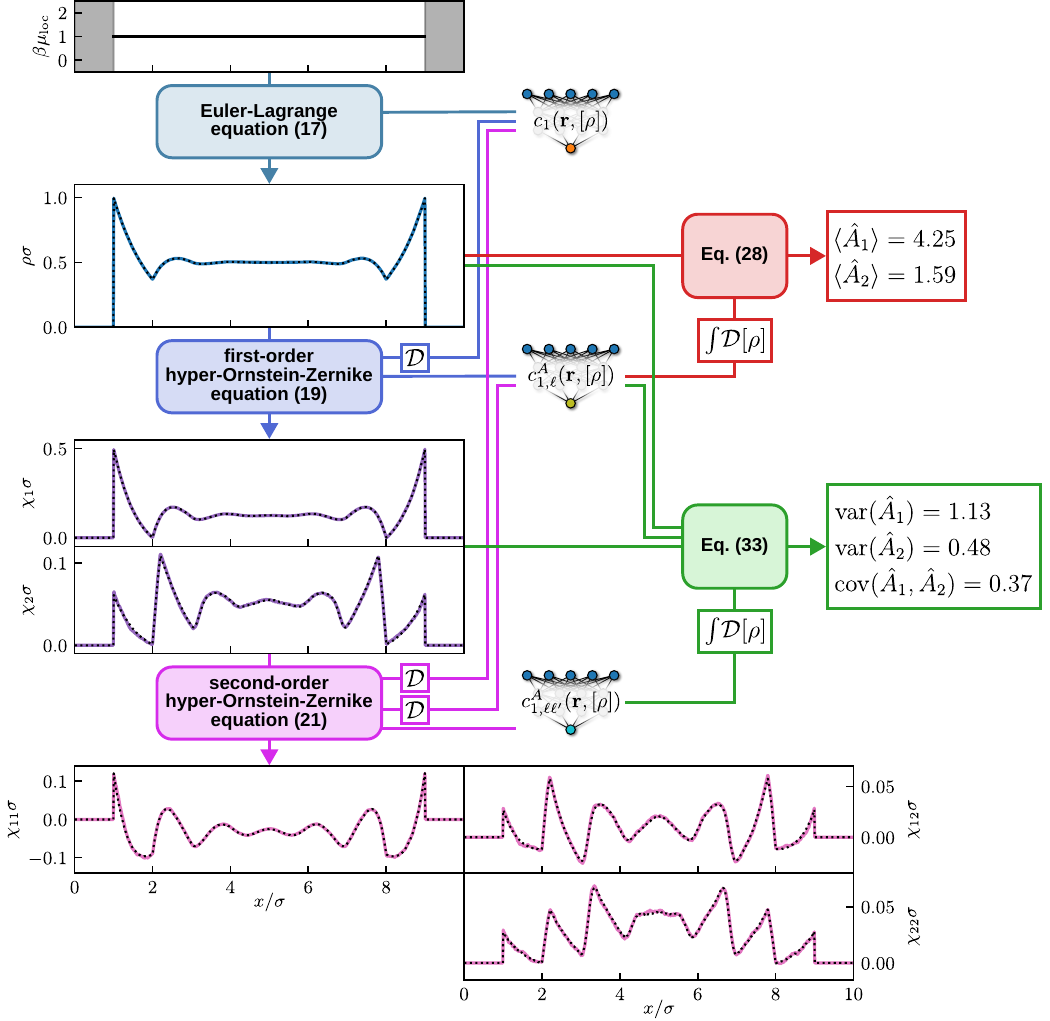}
  \caption{
    Prediction workflow for evaluating averages and covariances of chosen hyperobservables using neural hyperdensity functionals.
    As a prototypical case, we depict the one-dimensional ($\rv = x$) hard rod fluid under hard-wall confinement with walls at positions $x / \sigma = 1$ and $x / \sigma = 9$.
    The chosen hyperobservables are the total number of particles, $\hat A_1$, and the number of particles in the largest cluster, $\hat A_2$, see text.
    Via standard solution of the Euler-Lagrange equation~\eqref{eq:EulerLagrange} with the neural functional $c_1(\rv, [\rho])$, the density profile $\rho(x)$ corresponding to the imposed (scaled) local chemical potential $\beta \mu_\mathrm{loc}(x)$ is obtained.
    The associated first-order hyperfluctuation profiles $\chi_\ell(x)$, where $\ell = 1, 2$ for the depicted case, follow from solution of the first-order hyper-Ornstein-Zernike equation~\eqref{eq:hyperOZ}, where the neural hyperdirect correlation functional $c^A_{1,\ell}(\rv, [\rho])$ is invoked and $c_2(\rv, \rv', [\rho])$ is obtained from automatic differentiation of the neural functional $c_1(\rv, [\rho])$, as indicated by the symbol $\mathcal{D}$.
    Using functional line integration ($\int \mathcal{D}[\rho]$) of $c^A_{1,\ell}(\rv, [\rho])$ via Eq.~\eqref{eq:AellAsFunctionalIntegral}, the averages $A_\ell = \langle \hat A_\ell \rangle$ are evaluated.
    Eq.~\eqref{eq:AellellsAsFunctionalIntegral} yields the variances $A_{\ell\ell} = \var(\hat A_\ell)$ and covariances $A_{\ell\ells} = \cov(\hat A_\ell, \hat A_\ells)$, where $\ell, \ells = 1, 2$, via functional line integration of $c^A_{1,\ell\ells}(\rv, [\rho])$ together with the evaluation of the integrals in Eq.~\eqref{eq:AellellsAsFunctionalIntegral} incorporating $\chi_\ells(\rv)$ and $c^A_{1,\ell}(\rv, [\rho])$.
    Additionally, the second-order hyperfluctuation profiles $\chi_{\ell\ells}(x)$, are obtained from solution of the second-order hyper-Ornstein-Zernike equation~\eqref{eq:hyperOZsecondOrder}, where integral terms that do not involve $\chi_{\ell\ells}(x)$ are evaluated using directional functional derivatives, see Eq.~\eqref{eq:hyperOZsecondOrderGateaux}.
    The neural predictions for all one-body profiles match simulation results (black dotted lines), which serve as independent reference.
  }
  \label{fig:prediction_example}
\end{figure*}

\begin{figure*}
  \includegraphics{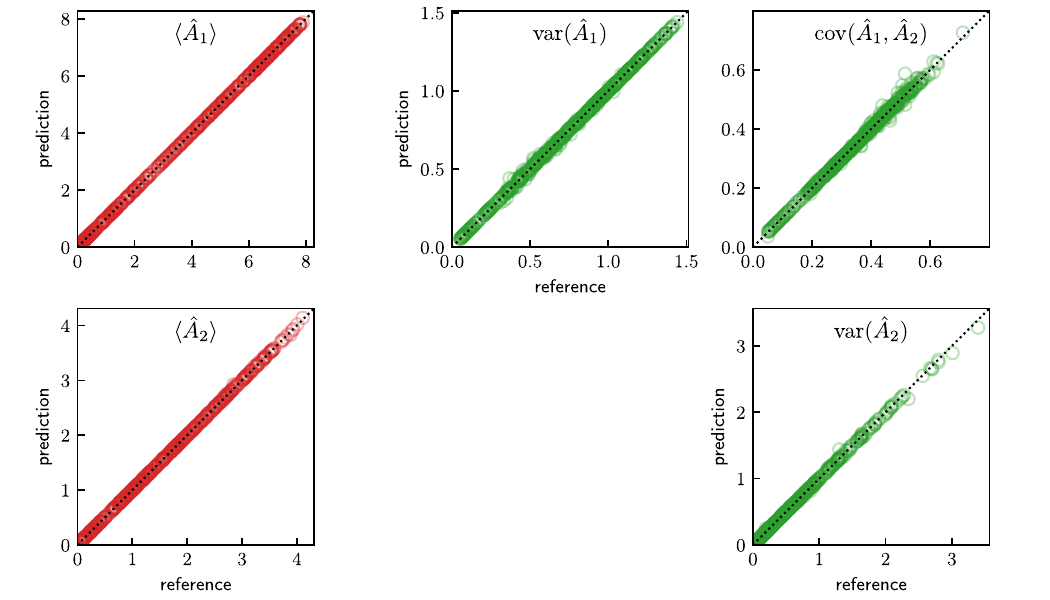}
  \caption{
    Predictions for a test data set of 512 simulations that have not been included in the training; see Fig.~\ref{fig:overview} for an illustration of the underlying machine learning protocol and Fig.~\ref{fig:prediction_example} for the prediction scheme.
    As before, the chosen hyperobservables are the total number of particles, $\hat A_1$, and the number of particles in the largest cluster, $\hat A_2$, see text.
    All relevant averages $\langle \hat A_1 \rangle$ and $\langle \hat A_2 \rangle$, variances $\var(\hat A_1)$ and $\var(\hat A_2)$, and the covariance $\cov(\hat A_1, \hat A_2)$ are recovered accurately across the entire test data set via the neural multivariate hyperdensity functional framework.
  }
  \label{fig:predictions}
\end{figure*}

For a practical application of our framework, we consider the one-dimensional hard rod fluid and wish to investigate its cluster statistics.
We recall that the Percus functional \cite{PercusEquilibriumStateClassical1976} permits exact analytic treatment in standard density functional theory and that neural functionals provide highly accurate surrogates \cite{SammullerWhyNeuralFunctionals2024,SammullerDeterminingChemicalPotential2025} in numerical work.
As shown in Refs.~\cite{SammullerHyperdensityFunctionalTheory2024,SammullerWhyHyperdensityFunctionals2025}, univariate (first-order) hyperdensity functional theory gives access to a single complex observable of choice in the form of a tractable neural hyperdensity functional.

We here consider the following multivariate case.
As the first hyperobservable, we choose the total number of particles within the system, $\hat A_1 = N$, which we recall is a fluctuating quantity in the grand ensemble.
For the second hyperobservable $\hat A_2$, the size of the largest cluster is chosen, which is an order parameter commonly used to investigate gel-forming systems \cite{SawComputerSimulationStudy2011,LindquistFormationEquilibriumGels2016,SammullerInhomogeneousSteadyShear2023}.
Two particles $i$ and $j$ are thereby defined as bonded if their distance is within a predefined cutoff, $|x_i - x_j| < x_c$, where $x_i$ and $x_j$ are the positions of the particles.
The cutoff distance $x_c = 1.2 \sigma$ is chosen in the following, where $\sigma$ is the particle diameter.
A cluster comprises of particles that are either directly or transitively bonded, and $\hat A_2$ is the number of particles in the largest cluster for a given microstate $x^N = x_1, \dots, x_N$.
The calculation of $\hat A_2$, which is hence indeed a (phase space) function of the configuration $x^N$, is performed with a clustering algorithm that remains applicable for arbitrary spatial dimensionality; code is provided in Ref.~\cite{GithubMultiHyper}.
We are ultimately interested in making the averages $A_1[\rho]$ and $A_2[\rho]$, variances $A_{11}[\rho]$ and $A_{22}[\rho]$, as well as the covariance $A_{12}[\rho] = A_{21}[\rho]$ explicitly accessible via neural hyperdensity functionals.

In order to generate a suitable training data set, we employ grand canonical Monte Carlo simulations and sample all of the required one-body profiles $\rho(\rv)$, $\chi_1(\rv)$, $\chi_2(\rv)$, $\chi_{11}(\rv)$, $\chi_{22}(\rv)$, and $\chi_{12}(\rv) = \chi_{21}(\rv)$ using standard histogram-counting methods.
The first- and second-order hyperfluctuation profiles are respectively calculated via the covariance and trivariance expressions \eqref{eq:chiellAsCovariance} and \eqref{eq:chiellellsAsTrivariance}, for which all required averages are sampled.
In total, 512 individual simulations are carried out at randomized scaled chemical potentials chosen uniformly in the range $-5.0 < \beta \mu < 5.0$.
In each simulation, a randomly generated external potential is imposed, which consists of hard walls, linear segments, and Fourier modes.
Typical external potential landscapes and corresponding results for a selection of pertinent one-body profiles are shown in Fig.~\ref{fig:overview}.
The machine learning of all neural (hyper)density functionals proceeds via the multistage protocol laid out in Sec.~\ref{sec:SimulationPerspective} and illustrated in Fig.~\ref{fig:overview}.
At each stage, the corresponding (hyper)direct correlation function $c_1(\rv)$, $c^A_{1,\ell}(\rv)$, or $c^A_{1,\ell\ells}(\rv)$ is calculated via post-processing of simulation data, thereby utilizing the availability of neural functionals of the previous stage.
For each of the thus obtained correlation functions, a neural network is trained to represent locally \cite{SammullerNeuralFunctionalTheory2023,SammullerWhyNeuralFunctionals2024,SammullerHyperdensityFunctionalTheory2024} the underlying functional dependence on the density profile.

The training of all neural functionals proceeds successfully, as is expected on the basis of the formal existence and uniqueness of the involved functional mappings.
We find consistently and reproducibly that the mean squared error loss function that we employ is minimized effectively upon standard optimization based on the underlying data set.
We take this as data-driven verification of the existence of all universal functionals that we seek to extract from the simulation results.

Employing the neural functionals for predicting $A_\ell[\rho]$ and $A_{\ell\ells}[\rho]$ via Eqs.~\eqref{eq:AellAsFunctionalIntegral} and \eqref{eq:AellellsAsFunctionalIntegral} yields values for the respective averages and covariances that closely match the simulation reference.
Besides the results for the exemplary case of hard-wall confinement reported in Fig.~\ref{fig:prediction_example}, we test in Fig.~\ref{fig:predictions} the neural predictions for a randomized data set of 512 simulations that have not been included in the training.
The agreement of all considered averages and covariances is very satisfactory across the entire data set.
Additionally, we demonstrate in Fig.~\ref{fig:prediction_example} the accessibility of the complete set of first- and second-order hyperfluctuation profiles, which are again validated against reference simulation data.
In conclusion, all quantities of interest can be predicted consistently with neural (hyper)density functionals; we recall Appendix~\ref{appendix:cov} for a numerical validation of equivalent routes for calculating covariances.
In line with previous work on standard neural density functionals \cite{SammullerNeuralFunctionalTheory2023,SammullerNeuralDensityFunctional2025}, the range of applicability thereby far exceeds the underlying simulation data and allows for extrapolation to unseen scenarios, such as the exemplary case of pure hard-wall confinement shown in Fig.~\ref{fig:prediction_example}, which has not been featured in the randomized training data set.

The specific choice of hyperobservables permits further identifications and consistency checks involving the arising hyperfluctuation profiles.
As $\hat A_1 = N$, the first-order hyperfluctuation profile $\chi_1(\rv)$ attains corresponding physical meaning as the (scaled) local compressibility $\kB T \chi_\mu(\rv)$; a well-known quantity used in previous work \cite{EvansLocalCompressibilityLiquids2015,EvansUnifiedDescriptionHydrophilic2019,WildingWhatBestSimulation2024,EckertFluctuationProfilesInhomogeneous2020,EckertLocalMeasuresFluctuations2023} to rationalize particle number fluctuations occurring in particular at substrates or solutes and close to phase transitions.
Returning to the exemplary case of hard-wall confinement (see Fig.~\ref{fig:prediction_example}), $\chi_1(\rv)$ shows typical enhanced fluctuations in the vicinity of the walls and oscillations at increasing distance from the walls follow closely the shape of the density profile.
The hard-wall contact value is $\chi_1(0^+) = \rho_b$ \cite{EckertLocalMeasuresFluctuations2023,SammullerWhyHyperdensityFunctionals2025}, where $\rho_b$ is the density of the bulk fluid far away from the wall.
In the present system, the bulk density may be approximated by the plateau value of $\rho(\rv)$ in the center of the box, thus verifying the contact theorem numerically.
Choosing the total number of particles $N$ as a hyperobservable constitutes a special case, as the corresponding hyperdirect correlation functional reduces analytically to $c^A_{1}(\rv) = 1$ \cite{EckertFluctuationProfilesInhomogeneous2020,EckertLocalMeasuresFluctuations2023,SammullerHyperdensityFunctionalTheory2024,SammullerWhyHyperdensityFunctionals2025}, which allows treatment already with standard (neural) density functional methods \cite{SammullerNeuralDensityFunctional2025}; we elaborate on further consequences below.

The behavior of $\chi_2(\rv)$, which localizes fluctuations of the size of the largest cluster, is remarkably different to that of $\chi_1(\rv)$.
The largest values of $\chi_2(\rv)$ are encountered away from the wall, at a distance approximately equal to the pairwise cluster cutoff criterion $x_c = 1.2 \sigma$, indicating that (i) a particle being present at this location is favorable for the occurrence of large clusters in the system and (ii) microscopic details of the chosen hyperobservable are deeply imprinted in the corresponding hyperfluctuation profile.
The hypercontact theorem $\chi_2(0^+) = A^b_2 / V$ \cite{SammullerWhyHyperdensityFunctionals2025} connects the hard-wall contact value $\chi_2(0^+)$ to the mean size of the largest cluster per volume, $A^b_2 / V$, in the bulk fluid.

The predicted second-order hyperfluctuation profiles $\chi_{11}(\rv)$, $\chi_{22}(\rv)$, and $\chi_{12}(\rv) = \chi_{21}(\rv)$ generally show more pronounced and rapid oscillations, as is consistent with the increased correlation level that is probed.
An interpretation of these quantities can be based on the trivariance definition~\eqref{eq:chiellellsAsTrivariance}.
For given hyperobservables $\hat A_\ell$ and $\hat A_\ells$, the second-order hyperfluctuation profile $\chi_{\ell\ells}(\rv)$ measures how a particle being present at location $\rv$ influences thermal fluctuations of the \emph{combination} of observables $\hat A_\ell \hat A_\ells$.
The specific form of the trivariance, cf.\ Eq.~\eqref{eq:triv}, thereby ensures that effects of mere covariances of $\hat A_\ell$ and $\hat A_\ells$ are excluded, as these are already captured by the first-order hyperfluctuation profiles.
A similar strategy for ``distilling'' higher-order fluctuation has been considered by \citeauthor{WildingCriticalSurfacePhase2025} \cite{WildingCriticalSurfacePhase2025}, who argue that doing so is favorable for capturing near-critical surface phase behavior in their solute-solvent system.
For our present scenario, one may glean from the results shown in Fig.~\ref{fig:prediction_example} that particle number-number fluctuations decrease toward the middle and that they are negative, i.e.\ anticorrelated, in large parts of the system.
Incorporating the largest cluster size changes this behavior, with $\chi_{22}(\rv)$ being positive everywhere and tending to increase toward the middle of the system.
Sharp maxima arise for $\chi_{22}(\rv)$ at discrete multiples of the cluster cutoff distance $x_c$ from the wall.
This demonstrates the capability of the second-order hyperfluctuation profiles to resolve increasing levels of \emph{microscopic} structuring while remaining firmly grounded on the one-body level.

Spatial integration of the density operator over the entire system domain yields $\int d\rv \hat\rho(\rv) = N$ by definition.
Besides the trivial relation $\int d\rv \rho(\rv) = \langle N \rangle$, the integrated hyperfluctuation profiles may therefore be reduced to:
\begin{align}
  \label{eq:chiellIntegral}
  \int d\rv \chi_\ell(\rv) &= \cov(N, \hat A_\ell),\\
  \label{eq:chiellellsIntegral}
  \int d\rv \chi_{\ell\ells}(\rv) &= \triv(N, \hat A_\ell, \hat A_\ells),
\end{align}
which follows from using their covariance and trivariance definitions~\eqref{eq:chiellAsCovariance} and~\eqref{eq:chiellellsAsTrivariance}.
Hence, when choosing $\hat A_1 = N$, Eq.~\eqref{eq:chiellIntegral} implies the identity $\int d\rv \chi_1(\rv) = A_{11}$.
Despite the apparent redundancy, the relation provides a valuable consistency check for the present conceptual investigation; indeed we find the sum rule to be verified numerically.
Note also that Eq.~\eqref{eq:chiellellsIntegral} yields $\int d\rv \chi_{11}(\rv) = \triv(N, N, N) = \langle N^3 \rangle - \langle N \rangle^3 - 3 \var(N) \langle N \rangle$, which allows one to quantify third-order particle number fluctuations in the present context.
We emphasize that the level of redundancy and the arising simplifications depend in general on the specific choice of hyperobservables, and that treating the total number of particles $N$ is already implicit in the theory by integration of hyperfluctuation profiles according to Eqs.~\eqref{eq:chiellIntegral} and \eqref{eq:chiellellsIntegral}.

\section{Conclusions}
\label{sec:Conclusions}

Based on the recent hyperdensity functional theory for soft matter systems \cite{SammullerHyperdensityFunctionalTheory2024,SammullerWhyHyperdensityFunctionals2025}, we have developed a first-principles-rooted approach to investigate the thermal equilibrium behavior of multiple variables that are simultaneously of interest.
The multivariate hyperdensity functional theory allows one to systematically study correlations and causal relationships of several order parameters in inhomogeneous fluids, soft matter, and general classical many-body systems featuring nontrivial internal and external interactions.
Our framework yields a concrete and practically feasible route toward expressing averages $A_\ell[\rho]$ and covariances $A_{\ell\ells}[\rho]$ of a set of general ``hyperobservables'' $\{\hat A_\ell\}$, $\ell = 1, \dots, L$, as universal density functionals.
Crucially, these encapsulate only intrinsic properties emerging from the interparticle interactions and they hence remain applicable to any specific physical situation as determined by the external environment.

The theory rests formally on the Mermin-Evans classical density functional mapping \cite{EvansNatureLiquidvapourInterface1979,EvansDensityFunctionalsTheory1992,HansenTheorySimpleLiquids2013}, whereby the fundamental minimization principle is considered in an extended ensemble that is induced by the set of hyperobservables $\{\hat A_\ell\}$.
Utilizing this extended ensemble determines uniquely first- and second-order hyperfluctuation profiles $\chi_\ell(\rv)$ and $\chi_{\ell\ells}(\rv)$ that are associated with the respective hyperobservables, cf.\ their definitions \eqref{eq:chiellAsDensityDerivative} and \eqref{eq:chiellellsAsDensityDerivative} as parametric derivatives of the density profile as well as the equivalent formulation in terms of the covariance \eqref{eq:chiellAsCovariance} and trivariance \eqref{eq:chiellellsAsTrivariance}.
The hyper-Ornstein-Zernike equations \eqref{eq:hyperOZ} and \eqref{eq:hyperOZsecondOrder} incorporate $\chi_\ell(\rv)$ and $\chi_{\ell\ells}(\rv)$ and they constitute defining integral equations for the first- and second-order hyperdirect correlation functionals $c^A_{1,\ell}(\rv,[\rho])$ and $c^A_{1,\ell\ells}(\rv,[\rho])$, which both possess universal density functional dependence.
Functional line integration of the hyperdirect correlation functionals, see Eqs.~\eqref{eq:AellAsFunctionalIntegral} and \eqref{eq:AellellsAsFunctionalIntegral}, respectively yields $A_\ell[\rho]$ and $A_{\ell\ells}[\rho]$ in density functional form.

We have substantiated the multivariate hyperdensity functional framework by considering the accessibility of all required functional mappings from both the simulation and machine learning perspective.
As is the case in standard classical density functional theory \cite{SammullerNeuralFunctionalTheory2023,SammullerWhyNeuralFunctionals2024,SammullerNeuralDensityFunctionals2024,SammullerNeuralDensityFunctional2025}, all functional relationships are amenable to be extracted from a simulation data set of inhomogeneous fluid equilibria and to be represented with neural networks.
In extension of the (univariate) first-order theory \cite{SammullerHyperdensityFunctionalTheory2024,SammullerWhyHyperdensityFunctionals2025}, the structure of the pertinent equations naturally gives rise to a multistage protocol for machine learning neural (hyper)direct correlation functionals, as illustrated in Fig.~\ref{fig:overview}.
The hyper-Ornstein-Zernike equations~\eqref{eq:hyperOZ} and \eqref{eq:hyperOZsecondOrder} lie at the heart of this methodology and a reformulation to their equivalent G\^{a}teaux forms~\eqref{eq:hyperOZGateaux} and \eqref{eq:hyperOZsecondOrderGateaux}, which feature (generalized) directional functional derivatives, turns out to be crucial for the numerical evaluation of all required expressions.
We have applied this workflow to the investigation of clustering statistics in the hard rod fluid, whereby the two hyperobservables $\hat A_1$ and $\hat A_2$ of interest were chosen respectively as the total number of particles in the system and the number of particles in the largest cluster, defined via a transitive proximity criterion.
Despite the simplicity of the system, the largest cluster size is an algorithmically defined many-body characteristic, making standard theoretical treatments arguably intractable in general.
Using neural hyperdensity functionals, all pertinent averages $A_1[\rho]$ and $A_2[\rho]$, variances $A_{11}[\rho]$ and $A_{22}[\rho]$, as well as the covariance $A_{12}[\rho]$ were shown to be explicitly accessible to high precision.
Additionally, all hyperfluctuation profiles $\chi_\ell(\rv)$ and $\chi_{\ell\ells}(\rv)$, $\ell, \ells = 1, 2$, could be determined accurately via solution of the hyper-Ornstein-Zernike equations \eqref{eq:hyperOZ} and \eqref{eq:hyperOZsecondOrder}, and these were shown to quantify systematically the localized fluctuations associated with the chosen hyperobservables.

We conclude with several conceptual points.

(i) The hyperdensity functional framework can be extended naturally to multicomponent systems that feature distinct particle species, as is relevant when dealing, e.g., with fluid mixtures \cite{RobitschkoLearningBulkInterfacial2025,MatthesGaugeInvarianceHyperforce2026}.
The species-labelling of all pertinent quantities and their inter- and cross-coupling via species-resolved hyper-Ornstein-Zernike equations is laid out in Appendix~\ref{appendix:Mixtures}.

(ii) Irrespective of the number of particle species, the theoretical derivations can be pursued beyond the second-order level.
Doing so gives rise to a hierarchy of higher-order hyper-Ornstein-Zernike equations that involve terms with increasing number of spatial integrals or, equivalently, with increasing derivative order.
However, in clear distinction to other well-known hierarchical equations in statistical mechanics, such as the BBGKY hierarchy~\cite{HansenTheorySimpleLiquids2013}, the resulting set of hyper-Ornstein-Zernike equations does not necessitate a closure relation at the level of truncation in order to be solvable.
Instead, all relevant expressions are explicitly accessible via neural functionals.
This ``inverted'' hierarchical structure arises due to starting from the functional Mermin-Evans minimization principle at the one-body level and applying derivatives instead of integrating out degrees of freedom in the full many-body phase space picture.

(iii) In a pedestrian approach, one could be tempted to circumvent the present second-order theory by treating all arising averages via first-order hyperdensity functional theory.
For instance, instead of determining the covariance $A_{\ell\ells} = \cov(\hat A_\ell, \hat A_\ells) = \langle \hat A_\ell \hat A_\ells \rangle - \langle \hat A_\ell \rangle \langle \hat A_\ells \rangle$ via the second-order route laid out in this work, one may in principle consider the averages $\langle \hat A_\ell \hat A_\ells \rangle$, $\langle \hat A_\ell \rangle$, and $\langle \hat A_\ells \rangle$ individually via the first-order theory \cite{SammullerHyperdensityFunctionalTheory2024,SammullerWhyHyperdensityFunctionals2025} and build the covariance a posteriori.
However, doing so has several severe drawbacks.
Numerically, the route is susceptible to catastrophic cancellation errors, as the averages that enter the covariance are generally much larger than the covariance itself.
Furthermore, the inherent theoretical structure that becomes apparent at the second-order level is entirely ignored.
We hence argue that treating covariances with the present second-order route, thereby incorporating physical constraints via the second-order hyper-Ornstein-Zernike equation in integral form~\eqref{eq:hyperOZsecondOrder} or in directional functional derivative form~\eqref{eq:hyperOZsecondOrderGateaux} and via the second-order hyperdirect correlation functional \eqref{eq:cA1ellells}, is both conceptually and practically beneficial as opposed to staying on the first-order level of the theory.

(iv) While we have kept our exemplary application in Sec.~\ref{sec:ClusterHardRods} simple regarding the considered geometry and fluid type, the theory and workflow that we have developed are general and remain applicable to more complex systems.
In Refs.~\cite{SammullerHyperdensityFunctionalTheory2024,SammullerWhyHyperdensityFunctionals2025}, the feasibility of machine learning univariate hyperdensity functionals has been demonstrated for the hard sphere fluid in planar three-dimensional geometry and we expect similar extensibility for the multivariate case.
In this regard, we emphasize that we deliberately employ (small) multilayer perceptrons for extracting and representing all (hyper)direct correlation functionals.
Despite their simple architecture, these neural networks already suffice in our applications as universal approximators of the required functional relationships.
However, we see much potential in tuning the neural network architecture to the application under consideration, in particular when dealing with complex systems and geometries, as becomes pertinent, e.g., for increased spatial resolution or when including orientational degrees of freedom.
Utilizing neural networks with built-in invariance or equivariance under specific symmetry operations may prove to be crucial for addressing these advanced applications, as exploited recently for constructing (standard) neural density functionals in two- and three-dimensional systems \cite{GlitschNeuralDensityFunctional2025,WeimarCubicEquivariantNeural2026,ChengEquivariantLearningTransferable2026}.

In future work, it may be interesting to consider a range of further theoretical developments and application scenarios.
It might be valuable to investigate the feasibility of deriving \emph{analytical} hyperdensity functionals.
This is clearly a significant challenge, albeit a well-posed one, whereby the availability of highly accurate neural hyperdensity functionals might serve as a useful aid and benchmark for finding workable analytic approximations.
Additional theoretical developments could encompass the inclusion of closure relations in the spirit of integral equation theories, which could serve as an alternative route to solving the hyper-Ornstein-Zernike equations.
Turning to practical applications, the systematic study of inhomogeneous fluids near solvents and substrates hinges on a thorough description of surface-induced physical mechanisms that encompass enhanced local fluctuations, lyo- and hydrophobicity, as well as wetting and drying transitions, which constitute pertinent problems to be tackled via the present theory.
\citeauthor{BuiFirstprinciplesApproachElectromechanics2025} \cite{BuiFirstprinciplesApproachElectromechanics2025} have laid out a variant of the hyperdensity functional framework to investigate charged and dielectric systems under spatial confinement \cite{BuiDielectrocapillarityExquisiteControl2025}.
Their recent work concerns the generalization to azeotropic mixtures \cite{ZhouRolesBulkSurface2026} and ab initio models of water \cite{BuiUnifiedMachineLearning2026}; it would be interesting to expand upon their findings via the inclusion of multiple species-resolved observables of interest.
Applications to further (complex) three-dimensional fluids featuring bespoke order parameters are a natural and highly challenging goal.
Pertinent examples include the investigation of colloidal equilibrium gels \cite{SawComputerSimulationStudy2011,SammullerInhomogeneousSteadyShear2023,LindquistFormationEquilibriumGels2016}, where the largest cluster size as investigated in this work is one of several central order parameters for quantifying the onset of gelation.
Further soft matter systems that necessitate the use of tailor-made order parameters for describing their emerging microstructure are two-dimensional systems and fluids with orientation-dependent interactions, where e.g.\ the hexatic order parameter arises as a key quantity.
One could also envisage \emph{designing} a set of suitable order parameters by analyzing multiple hyperobservables and their corresponding hyperfluctuation profiles in light of desired target properties.
Utilizing multivariate hyperdensity functional theory to deal with such intricate matters constitutes a highly valuable work program to be investigated in the future.

\section*{Data availability}

The data that support the findings of this article are openly available \cite{GithubMultiHyper}.

\begin{acknowledgments}
  We thank Silas Robitschko, Sophie Hermann, and Johanna Müller for useful discussions.
  This work is supported by the DFG (Deutsche Forschungsgemeinschaft) under Project No. 551294732.
\end{acknowledgments}

\appendix

\section{Comparison of covariance calculations}
\label{appendix:cov}

\begin{figure}[b]
  \includegraphics{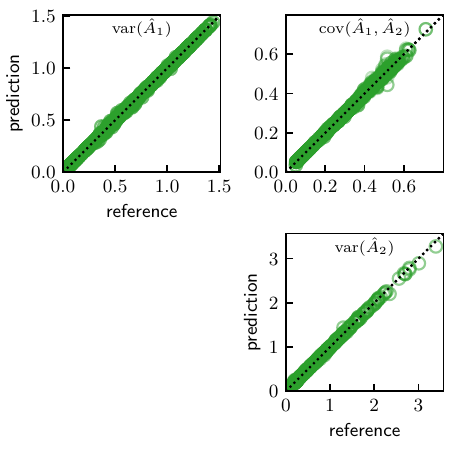}
  \includegraphics{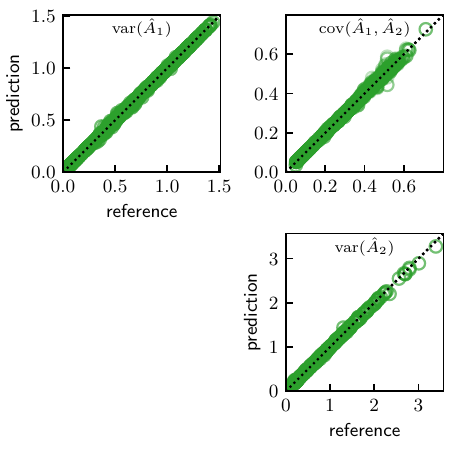}
  \caption{
    Results for variances and covariances evaluated via Eq.~\eqref{eq:AellellsAsFunctionalIntegral}, where $\ell$ and $\ells$ in the second term on the right-hand side is exchanged (top), and via Eq.~\eqref{eq:AellellsAsFunctionalIntegral2} (bottom).
    Both routes yield numerically equivalent results across the entire test data set, confirming the equivalence of both expressions and the exchange symmetry of the indices $\ell$ and $\ells$.
  }
  \label{fig:cov_comparison}
\end{figure}

Figure~\ref{fig:cov_comparison} shows a comparison of alternative implementations for calculating variances and covariances.
We evaluate Eq.~\eqref{eq:AellellsAsFunctionalIntegral} in its original form and with exchanged indices $\ell$ and $\ells$ in the second term on its right-hand side as well as Eq.~\eqref{eq:AellellsAsFunctionalIntegral2}.
All numerical results agree with each other, thus confirming the equivalence of the above expressions.

\section{Implementation of directional (G\^{a}teaux) derivatives via automatic differentiation in Julia}
\label{appendix:GateauxJulia}

\begin{figure*}[t]
\begin{lstlisting}[
  basicstyle=\small,
  label={listing:GateauxJulia},
  caption={Implementation of first-order and second-order directional (G\^{a}teaux) derivatives in the programming language Julia using the package ForwardDiff \cite{ForwardDiff.jl} for forward-mode automatic differentiation.}
]
using ForwardDiff: derivative

# First-order Gateaux derivative of F at rho in direction psi
function Gateaux(F, rho, psi)
    derivative(epsilon -> F(rho + epsilon * psi), 0)
end

# Second-order Gateaux derivative of F at rho in directions psi1 and psi2
function Gateaux(F, rho, psi1, psi2)
    derivative(epsilon -> Gateaux(F, rho + epsilon * psi2, psi1), 0)
end
\end{lstlisting}
\end{figure*}

Listing~\ref{listing:GateauxJulia} provides an implementation in the programming language Julia for calculating first- and second-order directional (G\^{a}teaux) derivatives generically and efficiently for a given functional $\mathcal{F}[\rho]$ using forward-mode automatic differentiation.

\section{Multivariate hyperdensity functional theory for mixtures}
\label{appendix:Mixtures}

For systems that are composed of several different components, much of the above theoretical structure persists and can be formulated straightforwardly, as we sketch in the following.
We label the different species by $\alpha=1,\dots, m$, where $m$ is the total number of components.
To facilitate book-keeping it is useful to introduce index sets ${\cal N}_\alpha$ that contain the labels $i$ of all particles that belong to species~$\alpha$.
The frequently occurring summation over all particles of a certain species~$\alpha$ can then be written compactly as $\sum_{i\in{\cal N}_\alpha}$.
An example that employs this convention is the partial density operator of species $\alpha$, written as $\hat\rho_\alpha(\rv)=\sum_{i\in{\cal N}_\alpha}\delta(\rv-\rv_i)$.

The standard mixture form of the Hamiltonian is $H=\sum_\alpha \sum_\ia \pv_i^2/(2m_\alpha) + u(\rv^N) +\sum_\alpha \sum_\ia V_\rmext^{(\alpha)}(\rv_i)$, where $u(\rv^N)$ is the interparticle interaction potential of the mixture, $V_\rmext^{(\alpha)}(\rv)$ is the external potential that acts on species $\alpha$ at position $\rv$, and $m_\alpha$ is the mass of particles of species $\alpha$; the sum $\sum_\alpha$ runs over all species.
The partition sum of the extended ensemble is $\Xi = \Tr \exp(-\beta(H-\sum_\alpha \mu_\alpha N_\alpha) + \sum_\ell \lambda_\ell \hat A_\ell)$, where $\mu_\alpha$ denotes the chemical potential and $N_\alpha$ the number of particles of species~$\alpha$.
The corresponding extended grand potential is then $\Omega=-\kB T \ln\Xi$.
Thermal averages in the extended ensemble are then built according to $\langle \cdot \rangle = \Tr \cdot \exp(-\beta(H-\sum_\alpha \mu_\alpha N_\alpha) + \sum_\ell \lambda_\ell \hat A_\ell)$.
An equivalent interpretation is that of an extended Hamiltonian $H_A = H - \sum_\ell \lambda_\ell \hat A_\ell/\beta$, which is coupled to particle baths, such that its Boltzmann factor is $\exp(-\beta(H_A - \sum_\alpha\mu_\alpha N_\alpha))$.
This leads to identical expressions for the partition sum and the grand potential as given above.

We follow the same structure as the one-component multivariate hyperdensity functional theory of Sec.~\ref{sec:Theory} in its following generalization to mixtures.
As in the one-component version, we consider a set $\{\hat A_\ell\}$, where the index $\ell=1,\dots,L$ enumerates the different observables $\hat A_\ell$, which can, but need not, be linked to specific species~$\alpha$.
Hence for generality, we keep the organization into species distinct from the choice of the observables.
In any practical application, there can of course be physically motivated interrelations.
As before, we aim at describing averages and covariances of the form \eqref{eq:AellDefinition} and \eqref{eq:AellellsDefinition}, respectively.

In generalization of Eqs.~\eqref{eq:AellAsParametricDerivative}--\eqref{eq:chiellellsAsDensityDerivative}, we obtain the following identities:
\begin{align}
  \label{eq:mixAellAsParametricDerivative}
  A_\ell &= -\frac{\partial\beta \Omega}{\partial\lambda_\ell} \Big|_\all{V_\rmext},\\
  \label{eq:mixAellellsAsParametricDerivative}
  A_{\ell\ells} &= -\frac{\partial^2 \beta \Omega}{\partial \lambda_\ell \partial \lambda_\ells} \Big|_\all{V_\rmext},
\end{align}
where all species-wise external potentials are kept fixed in the parametric differentiation, as indicated by the shorthand notation $\all{V_\rmext} = \{V_\rmext^{(\alpha)}(\rv)\}$.

The partial (i.e.\ species-resolved) first- and second-order hyperfluctuation profiles have the following form:
\begin{align}
  \chi_\ell^{(\alpha)}(\rv) &= \cov(\hat\rho_\alpha(\rv), \hat A_\ell),\\
  \label{eq:mixchiellellsViaTrivariance}
  \chi_{\ell\ells}^{(\alpha)}(\rv) &=\triv(\hat\rho_\alpha(\rv),\hat A_\ell, \hat A_\ells).
\end{align}

We recall that the partial density profile $\rho_\alpha(\rv) = \langle \hat{\rho}_\alpha(\rv) \rangle$ of species $\alpha$ can be obtained as the following functional derivative: $\rho_\alpha(\rv)= \delta \Omega / \delta V_\rmext^{(\alpha)}(\rv)$.
Similarly, functionally differentiating the mean $A_\ell$ and the covariance $A_{\ell\ells}$ by the (scaled) external potential for species~$\alpha$ yields:
\begin{align}
  \chi_\ell^{(\alpha)}(\rv) &= -\frac{\delta A_\ell}{\delta \beta V_\rmext^{(\alpha)}(\rv)},\\
  \chi_{\ell\ells}^{(\alpha)}(\rv) &= -\frac{\delta A_{\ell\ells}}{\delta \beta V_\rmext^{(\alpha)}(\rv)}.
\end{align}

Equivalently, the species-labelled first- and second-order hyperfluctuation profiles can be obtained as the following parametric derivatives of the density profile:
\begin{align}
  \chi_\ell^{(\alpha)}(\rv) &= \frac{\partial\rho_\alpha(\rv)}{\partial \lambda_\ell} \Big|_\all{V_\rmext},\\
  \chi_{\ell\ells}^{(\alpha)}(\rv) &= \frac{\partial^2\rho_\alpha(\rv)}{\partial\lambda_\ell\partial\lambda_\ells} \Big|_\all{V_\rmext}.
\end{align}

The Euler-Lagrange equations for the mixture are species-dependent and for species $\alpha$ given by:
\begin{align}
  \label{eq:elMixtureYetAgain}
  c_1^{(\alpha)}(\rv,[\all\rho]) &= \ln\rho_\alpha(\rv) + \beta V_\rmext^{(\alpha)}(\rv) - \beta\mu_\alpha,
\end{align}
where $c_1^{(\alpha)}(\rv,[\all\rho])$ is the one-body direct correlation functional for species $\alpha$, which depends functionally on the set of all partial density profiles, for which the shorthand notation $\all\rho = \{\rho_\alpha(\rv)\}$ is introduced.
The Euler-Lagrange equations \eqref{eq:elMixtureYetAgain} constitute a set of $m$ coupled equations, with one equation for each component~$\alpha$.
Parametrically differentiating Eq.~\eqref{eq:elMixtureYetAgain} with respect to $\lambda_\ell$ yields the following first-order hyper-Ornstein-Zernike relation:
\begin{equation}
  \label{eq:mixhyperOZ}
  \begin{split}
    &c_{1,\ell}^{A,(\alpha)}(\rv,[\all\rho]) =\\
    &\quad\frac{\chi_\ell^{(\alpha)}(\rv)}{\rho_\alpha(\rv)} - \sum_\alphas \int d\rv' \chi_\ell^{(\alphas)}(\rv') c_2^{(\alpha\alphas)}(\rv,\rv',[\all\rho]).
  \end{split}
\end{equation}
We have reordered the terms such that the right-hand side of Eq.~\eqref{eq:mixhyperOZ} contains all contributions that are accessible as averages in simulations, which encompasses the partial density profiles and $\chi_\ell^{(\alpha)}(\rv)$ via Eq.~\eqref{eq:mixchiellellsViaTrivariance}.
The species-resolved two-body direct correlation functionals may be obtained from analytical or neural functional representations.

The left hand side of Eq.~\eqref{eq:mixhyperOZ} is the (first-order) one-body hyperdirect correlation functional, as is specific to species $\alpha$ and observable $\ell$, defined as
\begin{equation}
  \label{eq:mixc1ell}
  c_{1,\ell}^{A,(\alpha)}(\rv,[\all\rho]) = \frac{\partial c_1^{(\alpha)}(\rv,[\all\rho])} {\partial \lambda_\ell} \Big|_\all\rho.
\end{equation}

The first-order two-body hyperdirect correlation functional is then defined as
\begin{align}
  c_{2,\ell}^{A,(\alpha\alphas)}(\rv, \rv', [\all\rho]) &= \frac{\partial c_2^{(\alpha\alphas)}(\rv,\rv',[\all\rho])} {\partial\lambda_\ell} \Big|_\all\rho\\
                                                        &=\frac{\delta c_{1,\ell}^{A,(\alpha)}(\rv,[\all\rho])} {\delta\rho_\alphas(\rv')},
\end{align}
where the latter version is obtained from considering the relationship between one- and two-body direct correlation functionals for a mixture: $c_2^{(\alpha\alphas)}(\rv,\rv',[\all\rho])= \delta c_1^{(\alpha)}(\rv,[\all\rho])/\delta \rho_\alphas(\rv')$.

Building the second parametric derivative of the species-resolved Euler-Lagrange equation \eqref{eq:elMixtureYetAgain}, or analogously differentiating the first-order hyper-Ornstein-Zernike relation \eqref{eq:mixhyperOZ} with respect to $\lambda_\ells$ gives the following second-order hyper-Ornstein-Zernike relation:
\begin{equation}
  \label{eq:mixhyperOZsecondOrder}
  \begin{split}
    &c_{1,\ell\ells}^{A,(\alpha)}(\rv,[\all\rho]) =\\
    &\quad\frac{\chi_{\ell\ells}^{(\alpha)}(\rv)}{\rho_\alpha(\rv)} -\frac{\chi_\ell^{(\alpha)}(\rv) \chi_\ells^{(\alpha)}(\rv)} {\rho_\alpha(\rv)^2}\\
    &\quad- \sum_\alphas\int d\rv' \chi_\ells^{(\alphas)}(\rv') c_{2,\ell}^{A,(\alpha\alphas)}(\rv,\rv',[\all\rho])\\
    &\quad- \sum_\alphas\int d\rv' \chi_\ell^{(\alphas)}(\rv') c_{2,\ells}^{A,(\alpha\alphas)}(\rv,\rv',[\all\rho])\\
    &\quad- \sum_{\alphas\alphass}\int d\rv'd\rv'' \chi_\ell^{(\alphas)}(\rv') \chi_\ells^{(\alphass)}(\rv'')\\
    &\qquad\qquad \times c_3^{(\alpha\alphas\alphass)}(\rv,\rv',\rv'', [\all\rho])\\
    &\quad- \sum_\alphas \int d\rv' \chi_{\ell\ells}^{(\alphas)}(\rv') c_2^{(\alpha\alphas)}(\rv,\rv',[\all\rho]).\\
  \end{split}
\end{equation}
Despite the increase in complexity over the first-order version \eqref{eq:mixhyperOZ}, here again the expressions on the right-hand side are available via sampling in simulations and numerical neural or approximate analytical functional methods.
In analogy to Eq.~\eqref{eq:cA1ellells}, the second-order one-body hyperdirect correlation functional for species $\alpha$ is
\begin{align}
  \label{eq:mixc1ellells}
  c_{1,\ell\ells}^{A,(\alpha)}(\rv,[\all\rho]) &= \frac{\partial^2 c_1^{(\alpha)}(\rv,[\all\rho])} {\partial \lambda_\ell \partial \lambda_\ells} \Big|_\all\rho\\
  \label{eq:mixc1ellellsAlt}
                                               &= \frac{\partial c_{1,\ell}^{A,(\alpha)}(\rv,[\all\rho])} {\partial \lambda_\ells} \Big|_\all\rho.
\end{align}
where the form \eqref{eq:mixc1ellellsAlt} is obtained from inserting Eq.~\eqref{eq:mixc1ell} into Eq.~\eqref{eq:mixc1ellells}.

As in the one-component multivariate theory, we can functionally integrate in the space of partial densities to retrieve the mean of the $\ell$th observable,
\begin{align}
  \label{eq:mixAellAsFunctionalIntegral}
  A_\ell[\all\rho] = \sum_\alpha \int d\rv \rho_\alpha(\rv) \int_0^1 da c^{A,(\alpha)}_{1,\ell}(\rv,[\all{a\rho}]),
\end{align}
and the covariance of the $\ell$th and $\ells$th observable,
\begin{equation}
  \label{eq:mixAellellsAsFunctionalIntegral}
  \begin{split}
    A_{\ell\ells}[\all\rho] &= \sum_\alpha \int d\rv \rho_\alpha(\rv) \int_0^1 da c_{1,\ell\ells}^{A,(\alpha)}(\rv,[\all{a\rho}])\\
    &\quad+\sum_\alpha \int d\rv \chi_\ells^{(\alpha)}(\rv) c_{1,\ell}^{A,(\alpha)}(\rv,[\all\rho]),
  \end{split}
\end{equation}
with the second integral in Eq.~\eqref{eq:mixAellellsAsFunctionalIntegral} being equivalent for exchanged labels $\ell$ and $\ells$.
In the parameterized functional line integrals of both Eqs.~\eqref{eq:mixAellAsFunctionalIntegral} and \eqref{eq:mixAellellsAsFunctionalIntegral}, all partial density profiles are scaled simultaneously, such that $\all{a\rho} = \{a\rho_\alpha(\rv)\}$ with identical scaling factor $0 \leq a \leq 1$ for all species $\alpha = 1, \dots, m$.
Other parameterizations of the functional line integrals are nevertheless conceivable.

For completeness, Eq.~\eqref{eq:mixAellellsAsFunctionalIntegral} can be rewritten in the following form:
\begin{equation}
  \label{eq:mixAellellsAsFunctionalIntegral2}
  \begin{split}
    &A_{\ell\ells}[\all\rho] =\\
    &\quad\sum_\alpha \int d\rv \rho_\alpha(\rv) \int_0^1 da c_{1,\ell\ells}^{A,(\alpha)}(\rv,[\all{a\rho}])\\
    &\quad+ \sum_\alpha \int d\rv \frac{\chi_\ell^{(\alpha)}(\rv)\chi_\ells^{(\alpha)}(\rv)} {\rho_\alpha(\rv)}\\
    &\quad- \sum_{\alpha\alphas} \int d\rv d\rv' \chi_\ell^{(\alpha)}(\rv) \chi_\ells^{(\alphas)}(\rv') c_2^{(\alpha\alphas)}(\rv,\rv',[\all\rho]),
  \end{split}
\end{equation}
analogously to Eq.~\eqref{eq:AellellsAsFunctionalIntegral2}.

The reformulation of integrals in the species-resolved hyper-Ornstein-Zernike equations~\eqref{eq:mixhyperOZ} and \eqref{eq:mixhyperOZsecondOrder} in terms of directional functional derivatives remains feasible by defining the species-aware G\^{a}teaux differential as
\begin{equation}
  \label{eq:mixGateaux}
  \begin{split}
    \mathcal{D}_\all\psi \mathcal{F}[\all\rho] &= \sum_\alpha \int d\rv' \psi_\alpha(\rv') \frac{\delta \mathcal{F}[\all\rho]}{\delta \rho_\alpha(\rv')}\\
                                     &= \frac{d \mathcal{F}[\{\rho_\alpha + \epsilon \psi_\alpha\}]}{d \epsilon} \Big|_{\epsilon = 0}
  \end{split}
\end{equation}
for a generic functional $\mathcal{F}[\all\rho]$, which may now depend in general on the entire set of partial density profiles $\all\rho = \{\rho_\alpha(\rv)\}$.
The directional functional derivative is then induced by the set $\all\psi = \{\psi_\alpha(\rv)\}$ of (generic) species-wise functions, as made explicit in the second line of Eq.~\eqref{eq:mixGateaux}.

Using Eq.~\eqref{eq:mixGateaux}, both multispecies hyper-Ornstein-Zernike equations~\eqref{eq:mixhyperOZ} and \eqref{eq:mixhyperOZsecondOrder} attain respectively their following compact G\^{a}teaux forms:
\begin{align}
  \label{eq:mixhyperOZGateaux}
  c^{A,(\alpha)}_{1,\ell}(\rv,[\all\rho]) &= \frac{\chi^{(\alpha)}_\ell(\rv)}{\rho_\alpha(\rv)} - \mathcal{D}_\all{\chi_\ell} c_1^{(\alpha)}(\rv, [\all\rho]),\\
  \label{eq:mixhyperOZsecondOrderGateaux}
  \begin{split}
    c^{A,(\alpha)}_{1,\ell\ells}(\rv,[\all\rho]) &= \frac{\chi^{(\alpha)}_{\ell\ells}(\rv)}{\rho_\alpha(\rv)} - \frac{\chi^{(\alpha)}_\ell(\rv)\chi^{(\alpha)}_\ells(\rv)}{\rho_\alpha(\rv)^2}\\
                                                 &\quad-\mathcal{D}_\all{\chi_\ells} c^{A,(\alpha)}_{1,\ell}(\rv, [\all\rho])\\
                                                 &\quad-\mathcal{D}_\all{\chi_\ell} c^{A,(\alpha)}_{1,\ells}(\rv, [\all\rho])\\
                                                 &\quad-\mathcal{D}_\all{\chi_\ell} \mathcal{D}_\all{\chi_\ells} c^{(\alpha)}_1(\rv, [\all\rho])\\
                                                 &\quad-\mathcal{D}_\all{\chi_{\ell\ells}} c^{(\alpha)}_1(\rv, [\all\rho]),
  \end{split}
\end{align}
in formal analogy to the single-species versions~\eqref{eq:hyperOZGateaux} and \eqref{eq:hyperOZsecondOrderGateaux}.

\bibliography{bibliography.bib}

\end{document}